\documentclass[longauth]{aa}  

\usepackage{graphicx}
\usepackage{txfonts}
\usepackage{amssymb,amsmath}
\usepackage[version=3]{mhchem}
\usepackage{threeparttable, tablefootnote}
\usepackage[section]{placeins}
\usepackage{hyperref}
\hypersetup{colorlinks = true,linkcolor= blue}
\usepackage{chemformula}
\usepackage[version=3]{mhchem}
\usepackage{lipsum}
\usepackage{stfloats}
\usepackage{amsmath}
\usepackage{nccmath}
\hypersetup{%
citecolor=blue
}
\usepackage[section]{placeins}
\usepackage{lscape}

\begin{document} 

\title{JOYS+ study of solid state \ce{^{12}C}/\ce{^{13}C} isotope ratios in protostellar envelopes}
\subtitle{Observations of CO and \ce{CO_2} ice with JWST}

\author{N. G. C. Brunken\inst{1}, E. F. van Dishoeck\inst{1,2}, K. Slavicinska\inst{1,3}, V. J. M. le Gouellec\inst{4}, W. R. M. Rocha\inst{1,3}, L. Francis\inst{1}, L. Tychoniec\inst{1}, M. L. van Gelder\inst{1}, M. G. Navarro\inst{6}, A. C. A.  Boogert\inst{7},  P. J. Kavanagh\inst{8}, P. Nazari\inst{9}, T. Greene \inst{4}, M. E. Ressler\inst{5} and L. Majumdar\inst{10,11}}

\institute{Leiden Observatory, Leiden University, 2300 RA Leiden, The Netherlands \\
\email{brunken@strw.leidenuniv.nl} 
\and 
Max-Planck-Institut f\"{u}r Extraterrestrische Physik, Gie{\ss}enbachstra{\ss}e 1, 85748 Garching, Germany 
\and
Laboratory for Astrophysics, Leiden Observatory, Leiden University, PO Box 9513, 2300 RA Leiden, The Netherlands 
\and
NASA Ames Research Center, Space Science and Astrobiology Division M.S 245-6 Moffett Field, CA 94035, USA 
\and
Jet Propulsion Laboratory, California Institute of Technology, 4800 Oak Grove Drive, Pasadena, CA 91109, USA 
\and
INAF-Osservatorio Astronomico di Capodimonte, Salita Moiariello
16, 80131 Napoli, Italy 
\and 
Institute for Astronomy, University of Hawai’i at Manoa, 2680 Woodlawn Drive, Honolulu, HI 96822, USA 
\and
Department of Experimental Physics, Maynooth University, Maynooth, Co. Kildare, Ireland 
\and
European Southern Observatory, Karl-Schwarzschild-Strasse 2, 85748 Garching bei München, Germany 510 
\and
School of Earth and Planetary Sciences, National Institute of Science Education and Research, Jatni 752050, Odisha, India 
\and
Homi Bhabha National Institute, Training School Complex, Anushaktinagar, Mumbai 400094, India 
}

\titlerunning{}
\authorrunning{N.G.C. Brunken et al.}

\date{Received; Accepted}

\abstract
{The carbon isotope ratio is a powerful tool for studying the evolution of stellar systems due to its sensitivity to the local chemical environment. Recent detections of CO isotopologues in disks and exoplanet atmospheres revealed high variability in the isotope abundance, pointing towards significant fractionation in these systems. In order to fully understand the evolution of this quantity in stellar and planetary systems however, it is crucial to trace the isotope abundance from stellar nurseries to the time of planet formation. During the protostellar stage the multiple vibrational modes of \ce{CO_2} and CO ice, that peak in the near- and mid-infrared, provide a unique opportunity to examine the carbon isotope ratio in the solid state. Now with the sensitivity and wide spectral coverage of the \textit{James Webb Space Telescope}, the multiple weak and strong absorption features of \ce{CO_2} and CO have become accessible at high signal to noise in Solar-mass systems.}
{We aim to study the carbon isotope ratio during the protostellar stage by deriving column densities and ratios from the various absorption bands of \ce{CO_2} and CO ice, and comparing our results with ratios derived in other astronomical environments.}
{We quantify the \ce{^{12}CO_2}/\ce{^{13}CO_2} and the \ce{^{12}CO}/\ce{^{13}CO} isotope ratios in 17 class 0/I low mass protostars from the \ce{^{12}CO_2} \ce{\nu_1} + \ce{\nu_2} and 2\ce{\nu_1} + \ce{\nu_2} combination modes (2.70 \ce{\mu}m and 2.77 \ce{\mu}m), the \ce{^{12}CO_2} \ce{\nu_3} stretching mode (4.27 \ce{\mu}m), the \ce{^{13}CO_2} \ce{\nu_3} stretching mode (4.39 \ce{\mu}m), the \ce{^{12}CO_2} \ce{\nu_2} bending mode (15.2 \ce{\mu}m), the \ce{^{12}CO} 1-0 stretching mode (4.67 \ce{\mu}m) and the \ce{^{13}CO} 1-0 stretching mode (4.78 \ce{\mu}m) using JWST NIRSpec and MIRI observations. We also report a detection of the 2-0 overtone mode of \ce{^{12}CO} at  2.35 \ce{\mu}m.}
{The column densities and \ce{^{12}CO_2}/\ce{^{13}CO_2} ratios derived from the various \ce{CO_2} vibrational modes are in agreement within the reported uncertainties and we find mean ratios of 85 $\pm$ 23, 76 $\pm$ 12 and 97 $\pm$ 17 for the 2.70 \ce{\mu}m band, the 4.27 \ce{\mu}m band and the 15.2 \ce{\mu}m band, respectively. The main source of uncertainty on the derived values stem from the error on the band strengths, the observational errors are in comparison negligible. Variation of the \ce{^{12}CO_2}/\ce{^{13}CO_2} ratio is observed from source to source which indicates that there could be genuine differences in the chemical conditions of their envelopes. The \ce{^{12}CO}/\ce{^{13}CO} ratios derived from the 4.67 \ce{\mu}m bands are consistent, albeit elevated with respect to the \ce{^{12}CO_2}/\ce{^{13}CO_2} ratios and we find a mean ratio of 165 $\pm$ 52. }
{These findings indicate that ices leave the pre-stellar stage with elevated carbon isotope ratios relative to the overall values found in the interstellar medium and that fractionation becomes a significant factor during the later stages of star and planet formation.}


\maketitle


\section{Introduction}

The ingredients constituting interstellar material, stars and eventually planets originate in dense molecular clouds where atoms and small molecules are accreted from the gas-phase onto cold dust grains forming icy mantles \citep{herbst2009complex,caselli2012,  boogert2015observations}. The abundance and composition of these ice mantles are therefore a direct probe of the chemical complexity of the molecular cloud at the time of ice formation. 
Susceptible to changes in these chemical conditions are isotope ratios that have been well studied both in the gas-phase and solid state in multiple celestial bodies \citep[and references therein]{nomura2022}. Variability of the isotope abundance across different astronomical environments is therefore indicative of the chemical evolution of these systems and the physicochemical processes that the pristine material has been subjected to. 

The carbon isotope ratio in particular is an attractive contender for isotope chemistry studies for numerous reasons. First, carbonaceous ices such as \ce{^{12}CO} and \ce{^{12}CO_2} are abundant and readily detected in infrared observations \citep{graauw1996sws, whittet1998detection, gerakines1999infrared,keane2001, pontoppidan2003m,bergin2005spitzer, pontoppidan2008c2d,oberg2011spitzer,mcclure2023ice} and if the sensitivity allows, their weaker isotopologue bands \ce{^{13}CO} and \ce{^{13}CO_2} are also observable in the same spectral range, thus enabling carbon isotope analysis in the solid state \citep{boogert1999iso, boogert2002a,pontoppidan2003m, brunken2024}. Additionally, CO and \ce{CO_2} ice have multiple vibrational modes  across the infrared spectrum that facilitate comparison between ratios derived from the different absorption bands. 

CO is also the second most abundant gas-phase molecule after \ce{H_2} in the interstellar medium (ISM) and its isotopologues \ce{^{12}C^{16}O}, \ce{^{13}C^{16}O}, and \ce{^{12}C^{18}O} therefore have strong gas-phase molecular lines in the sub-mm and infrared making them ideal for gas-phase studies in various environments \citep{smith2015,nomura2022}, including exo-planet atmospheres \citep{zhang2021, zhang2021b,line2021,ghandi2023, regt2024}.  This plethora of observational possibilities enables us to draw comparisons between gas-phase and solid state abundances and further elucidate the origin of interstellar material and its evolution from stellar nurseries to the time of planet formation. 

Enrichment of interstellar material with \ce{^{13}C} initially occurs during the CNO-cycle of stellar nucleosynthesis ($M_{\star}$ > 1.3 $M_{\odot}$) when \ce{^{13}N} atoms are converted into \ce{^{13}C} atoms. A fraction of these \ce{^{13}C} atoms remain in the star and is later released into the ISM at the end of its life-cycle \citep{milam2005,milam2009, prantzos2008}.
Once formed, the main fractionation processes responsible for affecting regional carbon isotope ratios are isotope exchange reactions, isotope-selective photodissociation, and possibly gas-ice partitioning \citep{watson1976,langer1984,langerpenzias1993, dieshoeckblack1988,visser2009,smith2015}. 

The yardsticks against which all observations are usually measured are the ratios derived for the local ISM by \citet[\ce{^{12}C}/\ce{^{13}C} $\sim$ 77 $\pm$ 7]{wilsontrood1994} from CO studies and by \citet[\ce{^{12}C}/\ce{^{13}C} $\sim$ 68 $\pm$ 15]{milam2005} from CN studies as well as the Solar abundance ratio \citep[\ce{^{12}C}/\ce{^{13}C} $\sim$ 89]{wilsontrood1994}.
In diffuse clouds, \citet{ritchey2009} derived ratios from optical line absorption studies of \ce{CH^+} that were consistent with the ISM standard (\ce{^{12}CH^+}/\ce{^{13}CH^+} $\sim$ 74). Conversely, the effect of selective photodissociation was observed in other diffuse clouds where higher ratios were extracted from CO measurements (\citet[\ce{^{12}CO}/\ce{^{13}CO} $\sim$ 167]{lambert1994} and \citet[\ce{^{12}CO}/\ce{^{13}CO} $\sim$ 125 and 117]{federman2003}).

In dense star forming regions \citet{jorgensen2016,jorgensen2018} found mainly sub-ISM ratios from gas-phase studies of complex organic molecules (COMs) with considerable deviations between the different species (\ce{^{12}C}/\ce{^{13}C} $\sim$ 27-67). Large discrepancies were also observed in studies of protoplanetary disks where the values varied drastically depending on the line of sight and the region of the disk in which the ratio was measured \citep{ hily-blant2019, booth2019b,booth2024a, bergin2024}. Finally, cometary ratios and ratios derived from chondrites were found to be similar to the standard Solar value ($\sim$ 89) \citep{alexander2007,bockelee2015,hassig2017,nomura2022}. 

In the solid state, \citet{boogert1999iso} examined a number of massive young stellar objects (MYSOs), three low mass young stellar objects (LYSOs), and two clouds and found \ce{^{12}CO_2}/\ce{^{13}CO_2} $\sim$ 52 - 113, consistent with the standard ISM ratio. Additional studies of CO ice for one MYSO \citep{boogert2002b} and one LYSO \citep{pontoppidan2003m} yielded \ce{^{12}CO}/\ce{^{13}CO} $\sim$ 71 and 69, respectively. The sample of solar-mass protostars remained limited however since past observational facilities lacked the sensitivity required to observe most of these low mass objects. Moreover, oftentimes the partial spectral coverage of these observatories also meant that CO and \ce{CO_2} ice could not be examined simultaneously.  

Now with the exquisite sensitivity of the \textit{James Webb Space Telescope} (JWST) \citep{gardner2023,wright2023,argyriou2023, jakobsen2022,boker2022}, we are able to study the various vibrational modes of CO and \ce{CO_2}, including their weaker isotopologue absorption features, for the first time in solar-mass protostars. Moreover, the high spectral resolution of the JWST offers a unique opportunity to perform  detailed profile analysis of these strong and weak ice features. \citet{mcclure2023ice} for instance, utilized this unprecedented sensitivity to quantify solid state \ce{^{12}CO_2}/\ce{^{13}CO_2} and \ce{^{12}CO}/\ce{^{13}CO} ratios in the Cha I dark molecular cloud (\ce{^{12}CO_2}/\ce{^{13}CO_2} $\sim$ 69 - 87 and \ce{^{12}CO}/\ce{^{13}CO}$\sim$ 99 - 169). \citet{brunken2024} also used the high spectral resolution of the JWST to perform detailed profile analysis of the 4.39 \ce{\mu}m band of \ce{^{13}CO_2} and showed that the band can be decomposed in five principal components \citep{pontoppidan2008c2d} that are representative of the chemical and thermal environment of the \ce{CO_2} ice.

In this paper we present high resolution JWST NIRSpec and MIRI observations of the \ce{^{12}CO_2} \ce{\nu_1} + \ce{\nu_2} and the 2\ce{\nu_1} + \ce{\nu_2} combination modes (2.70 \ce{\mu}m and 2.77 \ce{\mu}m), the \ce{^{12}CO_2} \ce{\nu_3} stretching mode (4.27 \ce{\mu}m), the \ce{^{13}CO_2} \ce{\nu_3} stretching mode (4.39 \ce{\mu}m), the \ce{^{12}CO_2}  \ce{\nu_2} bending mode (15.2 \ce{\mu}m), the \ce{^{12}CO} 2-0 overtone mode (2.35 \ce{\mu}m), the \ce{^{12}CO} 1-0 stretching mode (4.67 \ce{\mu}m) and the \ce{^{13}CO} 1-0 stretching mode (4.78 \ce{\mu}m) for 17 low mass Class 0/I protostars observed as part of the JWST Observations of Young protoStars (JOYS+) program \citep[Ressler et al. in prep.]{dishoeck2023,beuther2023}. We derive column densities of \ce{^{12}CO_2}, \ce{^{13}CO_2}, \ce{^{12}CO} and \ce{^{13}CO} for each vibrational mode and determine the carbon isotope ratio for the sources in this sample. We build on the work of \citep{boogert1999iso} by expanding the sample of solar-mass objects and examining their isotope reservoir. 

This paper is structured as follows. In Section \ref{sec:2} we present our data and describe the data reduction process and the methods for analyzing the spectra. In Section \ref{sec:3} we derive column densities and isotope ratios for the bands as a whole, and we discuss the results in Section \ref{sec:4}. Finally, in Section \ref{sec:5} we provide a summary and the concluding remarks.

\section{Observations and Methods}
\label{sec:2}

\subsection{Observations}

\begin{figure*}[h!]
    \centering
     \includegraphics[width=0.9\hsize]{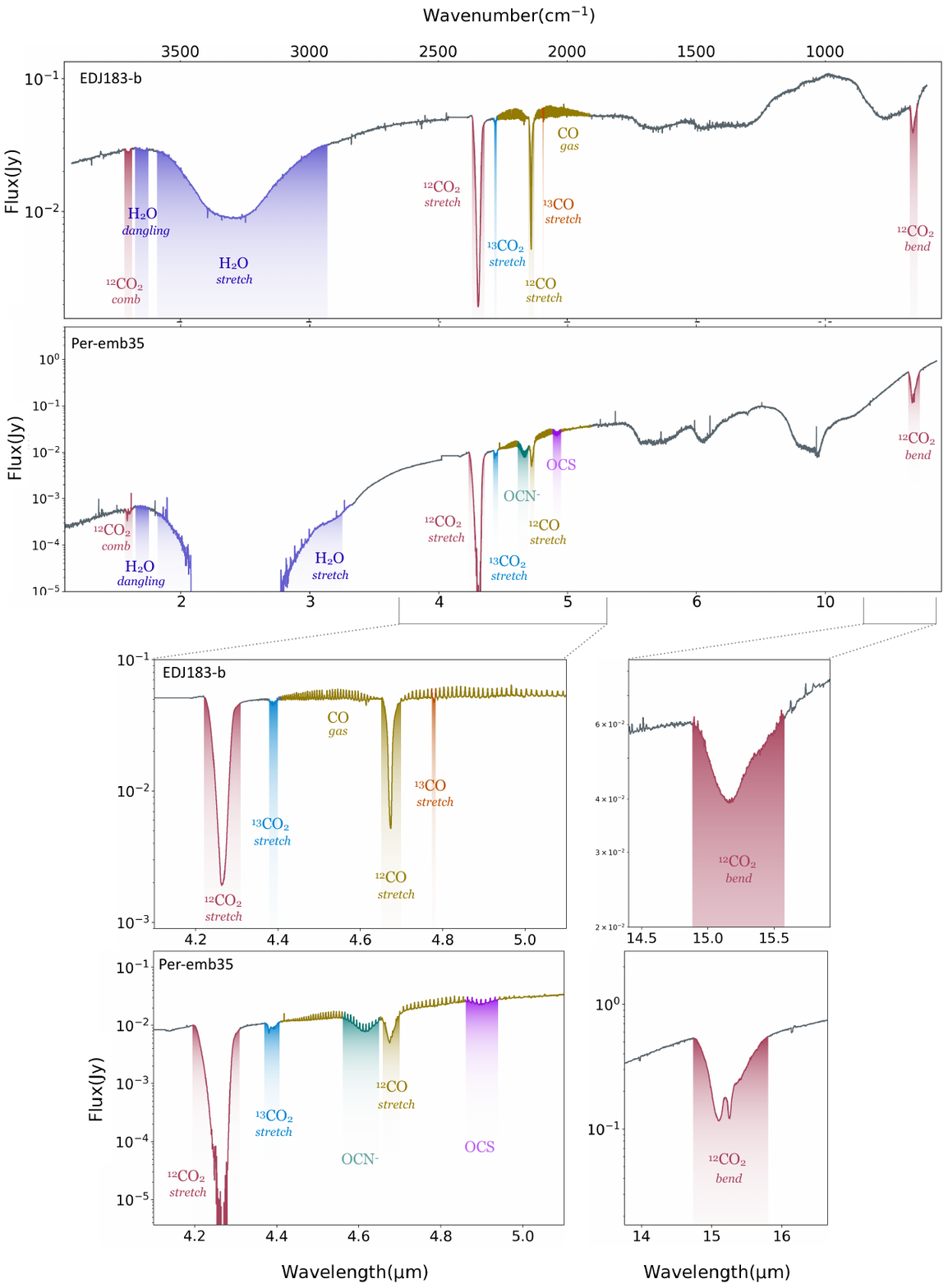}
    \caption{Full NIRSpec and MIRI spectra for EDJ183-b and Per-emb35. The the ice absorption features are shaded in color. The bottom panels show zoomed-in sections of the spectra.}
    \label{fig:overview}
\end{figure*}

Our observations were taken as part of the JWST Observations of Young protoStars (JOYS+) Cycle 1 Guaranteed Time Program (GTO) NIRSpec (PI: E.F. van Dishoeck, ID: 1960),  MIRI (PI: M. E. Ressler, ID: 1236). The data consist of NIRSpec and MIRI spectra of 17 Class 0/I sources including binaries observed using the G235M, G235H, G395M and G395H modes ($R$ = $\lambda$ /$\Delta$ $\lambda$ = 1000 and 2700 for the NIRSpec observations, respectively). 

The NIRSpec data cubes were reprocessed using the JWST pipeline (version 1.13.4) and the corresponding CRDS context jwst\_1231.pmap (CRDS\_VER = `11.17.19'). Additional corrections outside the JWST pipeline were applied to enhance the data quality. The calwebb\_detector1 step was executed with standard parameters. Subsequently, NSClean \citep{Rauscher2024} was utilized to address faint vertical banding and "picture frame noise" in the rate files. Following this, calwebb\_spec2 was run to generate the cal files. A systematic search was conducted in the cal files to identify warm pixels for flagging, as detected in the MAST final cubes. Warm pixels labeled as UNRELIABLE\_FLAT and NO\_SAT\_CHECK were flagged to prevent their usage in the final product production. The remaining not-flagged warm pixels were identified using sigma clipping with high sigma values to prevent the inadvertent clipping of real data. Subsequently, the DQ extension of the cal files was updated to incorporate these newly flagged pixels. Finally, calwebb\_spec3 was executed, configuring the outlier detection in the JWST outlier\_detection step with a threshold\_percent = 99.9 and a kernel\_size = 7 7. As a result, the final cubes exhibit improved cleanliness and a significantly higher signal-to-noise ratio compared to those obtained directly from the MAST.

For the MIRI cubes, a two-point dither pattern in the negative direction was performed for the majority of the sources on the source position. One exception is B1-c for which a four-point dither pattern was used. For each star-forming region (i.e., Taurus, Perseus, Serpens), one dedicated background was taken in a two-point dither pattern, except for in Taurus where a single dither was used, in order to properly subtract the telescope background and to better remove detector artifacts. All observations were carried out using the FASTR1 readout mode and use all three MIRI-MRS gratings (A, B, C), providing the full 4.9-28.6~$\mu$m wavelength coverage. 

The data were processed through all three stages of the JWST calibration pipeline version 1.13.4 \citep{Bushouse2023} using the same procedure as described by \citet{gelder2024}. The reference contexts {\tt jwst$\_$1188.pmap} of the JWST Calibration Reference Data System (CRDS; Greenfield \& Miller 2016) was used. In short, the raw data were processed through the {\tt Detercor1Pipeline} using the default setting. The dedicated backgrounds were subtracted on the detector level in the {\tt Spec2Pipeline}. This step also included the correction by the fringe flat for extended sources (Crouzet et al. in prep.) and a residual fringe correction (Kavanagh et al. in prep.). In order to remove any remaining warm and bad pixels from the calibrated detector data, an additional bad pixel routine was applied using the Vortex Imaging Processing (VIP) package \citep{Christiaens2024}. Lastly, the final datacubes were constructed for each band of each channel separately using the {\tt Spec3Pipeline}.

The properties of the Perseus objects studied in this work can be found in \citet{tobin2016}. An overview of the ice bands discussed in this work is presented in Figure \ref{fig:overview} for the two sources EDJ183-b and Per-emb35.

\subsection{Spectral extractions}

Spectral extractions were done for both the NIRSpec and MIRI range at central position using a cone diffraction extraction method with a 3$\lambda$/$D$ cone aperture. We opted for this aperture size to ensure that a maximum amount of flux was being included while avoiding overlap between the binary sources as much as possible. Overlap in the long wavelength MIRI channels was unavoidable due to the increasing size of the point spread function. The identifying names and spectral extraction coordinates are given for all sources in Table \ref{Tab:coordinates}. Binary systems are denoted with `a' and `b'.

\subsection{Continuum subtraction} 

Prior to the analysis the spectra are converted from flux scale (\textit{\ce{F^{\rm{obs}}_{\lambda}}}) to optical depth scale using equation \ref{eq:OD1}:

\begin{equation}
\label{eq:OD1}
    \tau^{\rm{obs}}_{\lambda} = -ln\biggl(\frac{F^{\rm{obs}}_{\lambda}}{F^{\rm{cont}}_{\lambda}}\biggl),
\end{equation}
where \textit{\ce{F^{\rm{cont}}_{\lambda}}} is the flux of the continuum. In the following sections we will discuss the continuum placement for each vibrational mode. 

\begin{figure*}[h!]
    \centering
     \includegraphics[width=0.9\hsize]{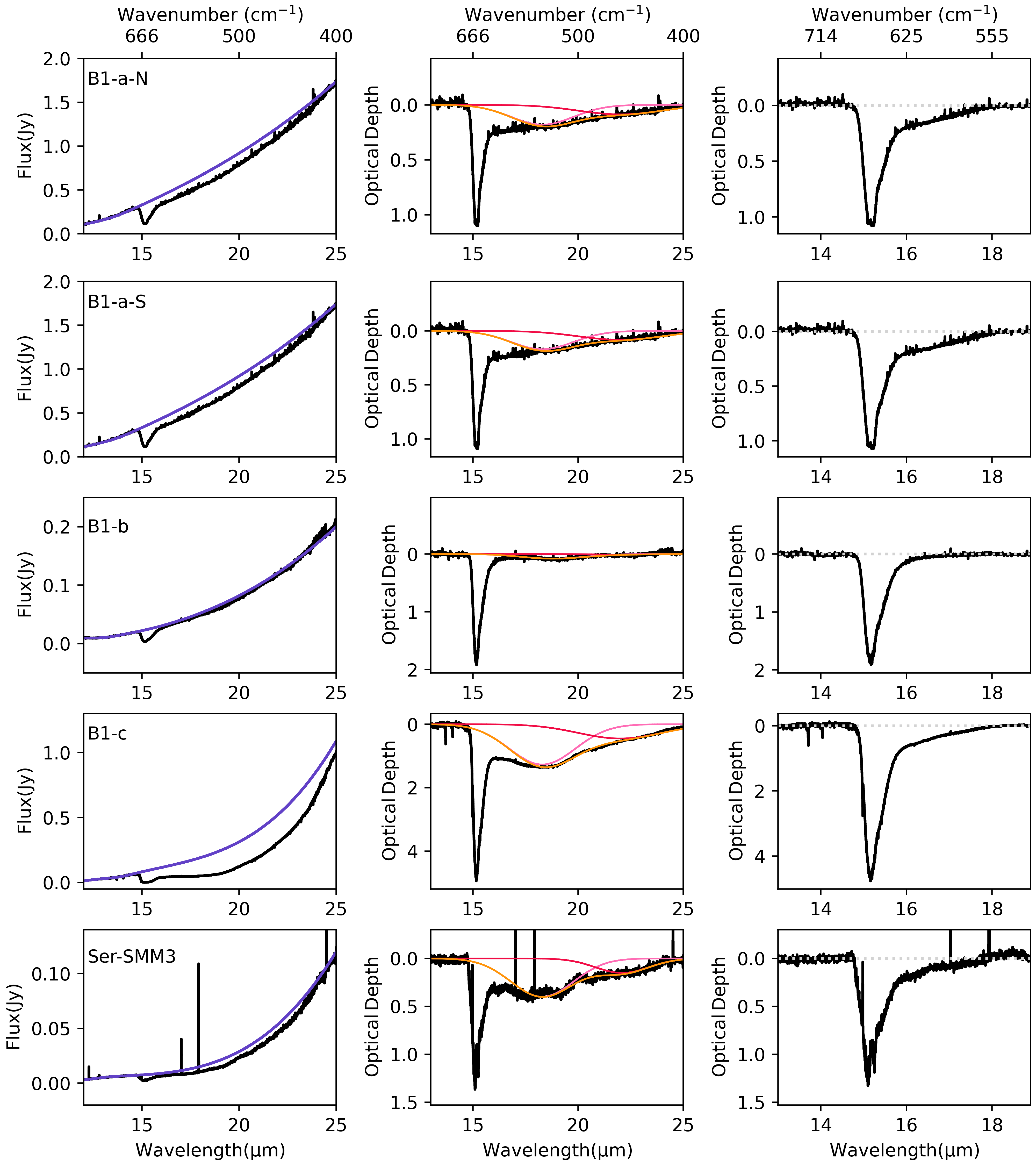}
    \caption{Continuum and silicate subtraction for the \ce{^{12}CO_2} bending mode in the 15 \ce{\mu}m region. The left panel shows the fitting of the continuum (purple line). The middle panel shows the two Gaussian profiles (red and pink lines) and the combined final fit (orange line) that simulates the silicate absorption feature at 18 \ce{\mu}m. The third panel shows the subtracted spectra after the silicate band removal. Continuum subtraction for the remaining sources can be found in Appendix \ref{app:MIRI-continuum}.}
    \label{fig:cont15micron1}
\end{figure*}

\subsubsection{The 2 \ce{\mu}m region}
\label{sec:2.1.1}

The NIRSpec G235M and G235H modes covering the CO and \ce{CO_2} bands in the 2 \ce{\mu}m region are available for 9 of the 17 sources. Local continuum subtractions were performed for the \ce{^{12}CO} overtone mode (2.35 \ce{\mu}m) and the \ce{^{12}CO_2} combination modes (2.70 \ce{\mu}m and 2.77 \ce{\mu}m) by fitting a third order polynomial to the data as illustrated in Figure \ref{fig:cont2micron}. For the overtone mode we selected points between 2.30 - 2.34 \ce{\mu}m and between 2.36 - 2.38 \ce{\mu}m while being mindful of the noise spikes and gas phase lines in this region. Local continua for the 2.70 \ce{\mu}m bands were drawn using a similar method as \citet{keane2001}.

\subsubsection{The 4 \ce{\mu}m region}
\label{sec:2.1.2}

The \ce{^{12}CO_2} stretching (4.27 \ce{\mu}m), the \ce{^{13}CO_2} stretching mode (4.39 \ce{\mu}m), the \ce{^{12}CO} stretching mode (4.67 \ce{\mu}m) and the \ce{^{13}CO} stretching mode (4.78 \ce{\mu}m) all have absorption features in the 4 \ce{\mu}m spectral region. For many sources this region is also heavily dominated by the CO gas-phase rotation-vibrational lines that in some cases even overlap with the \ce{^{13}CO_2} ice bands \citep{brunken2024}. For this region we performed a local continuum subtraction by selecting points between 4.30 - 4.36 \ce{\mu}m, 4.41 - 4.60 \ce{\mu}m, 4.71 - 4.80 \ce{\mu}m and 4.91 - 5 \ce{\mu}m. We opted to place a local continuum over the 4.30 - 4.9 \ce{\mu}m range since the shape of the continuum could be better traced once we expanded the wavelength range, especially for sources with a strong gas-line forest \citep{rubinstein2023,federman2023}. In Figure \ref{fig:cont4micron} we show the continuum fittings for two sources B1a-b (strong CO gas-line forest) and L1527 (weak CO gas-line forest). 

We performed separate local continuum subtractions for the \ce{^{12}CO_2} 4.27 \ce{\mu}m stretching mode due to the sensitivity of this band to scattering effects \citep{dartois2022influence, dartois2024}. Continuum points were placed between 4.17 - 4.19 \ce{\mu}m and 4.31 - 4.35 \ce{\mu}m following the methods described in \citet{dartois2022influence}. The results are presented in Figure \ref{fig:continuum-co2-str}.

\subsubsection{The 15 \ce{\mu}m region}
\label{sec:cont-15}

Continuum subtraction in the 15 \ce{\mu}m region is particularly challenging due to several deep absorption and, in some cases, emission features that collectively alter the shape of the continuum. Two significant features for instance are the silicate bending and stretching modes that have long been suspected of distorting the wing of the \ce{^{12}CO_2} band. The two strong absorption features are located at 9.7 and 18 \ce{\mu}m, with the latter likely pushing down the long-wavelength wing of the 15.2 \ce{\mu}m \ce{CO_2} band when the silicate is in absorption. This produces a band profile that is eerily similar to the grain shape effects observed for the \ce{CO_2} stretching mode (4.27 \ce{\mu}m) where the short-wavelength wing of the band is raised relative to the long-wavelength wing creating a `negative' absorption feature \citep{dartois2022influence,dartois2024}. 

Due to this striking resemblance we briefly considered the possibility that the distortion at 15.2 \ce{\mu}m is also the result of scattering effects but we argue that in order to produce such a strong scattering feature at these long wavelengths, grain sizes in the order of $\sim$10 \ce{\mu}m are required, which is an unlikely scenario in dense protostellar envelopes where the grain sizes are in the order of $\sim$ 0.9 \ce{\mu}m \citep{dartois2024}. Furthermore, for sources where the silicate band is in emission (Figure \ref{fig:silicate-emission}), we see the opposite effect where the short-wavelength wing of the \ce{CO_2} band is lowered with respect to the long-wavelength wing since the silicate band is pushing the long-wavelength wing `up' instead of `down'. In addition to the silicate absorption (and emission) band, water has a broad liberation mode that peaks at 13.6 \ce{\mu}m and extends over the 15 - 28 \ce{\mu}m region which could further contribute to the total absorption.

Taking all of these factors into account, we first placed a local continuum by selecting points between 12 - 14.5 \ce{\mu}m and 25 - 28 \ce{\mu}m and converted the spectra to optical depth scale (Figures \ref{fig:cont15micron1}, \ref{fig:cont15micron2}, \ref{fig:cont15micron3} and \ref{fig:cont15micron4}, first column). We subsequently modelled the asymmetric shape of the silicate bending mode with two Gaussian profiles centered at 18.3 \ce{\mu}m and 22 \ce{\mu}m respectively, and subtracted this from the optical depth spectra (Figures \ref{fig:cont15micron1}, \ref{fig:cont15micron2}, \ref{fig:cont15micron3} and \ref{fig:cont15micron4}, second column). After subtraction, the bands still bear an extended long-wavelength wing (Figures \ref{fig:cont15micron1}, \ref{fig:cont15micron2}, \ref{fig:cont15micron3} and \ref{fig:cont15micron4}, third column) but the short-wavelength wing is no longer significantly raised with respect to the long-wavelength wing. This method of modelling the silicate band with Gaussian profiles was also used by  \citet{gerakines1999infrared} and \citet{pontoppidan2008c2d}. For the two sources where the silicate band is in emission, we fitted an additional Gaussian profile centered at 17 \ce{\mu}m to simulate the shape of the silicate band. 

Finally, we investigated the contribution of the water liberation mode but found that the wing of this water band is almost identical to the local continuum we placed between the 12 - 28 \ce{\mu}m and that it therefore has no additional contribution once we subtract the continuum from the spectra.

\subsection{Column densities}

Column densities for the vibrational modes are derived using

\begin{equation}
\label{eq:CD1}
    N = \frac{\int\tau d\nu}{A},
\end{equation}

where $\tau_{\nu}$ is the integrated absorbance under the absorption feature and $A$ is the corresponding band strength of the vibrational mode. 

For the band strengths we used the values determined by \citet{gerakines1995} and \citet{gerakines2005} and corrected by \citet{bouilloud2015}. All band strengths used in this work are listed in Table \ref{Tab:bandstrengths1} and are further discussed along with the error analysis on the column densities in Appendix \ref{Appendix:A}. The column densities derived in this work can be multiplied by the correction factors given in Table \ref{Tab:bandstrengths1} to compare with values derived using the band strengths reported in \citet{gerakines1995} and \citet{gerakines1995}. We note that the recent band strengths \citet{gerakines2015} derived for amorphous \ce{CO_2} are similar to the corrected values we are using in this work. The change in the band strength due to the ice mixture and the temperature are accounted for in the error analysis. Lastly, the \ce{^{12}C}/\ce{^{13}C} ratios are determined for the bands as a whole and not for the individual components.

\section{Analysis}
\label{sec:3}

\begin{center}
\begin{table*}[hbt!]
\caption{Band strengths of \ce{CO_2} and CO ice.}
\small
\centering
\begin{tabular}{llccc}
\hline \hline
Position (\ce{\mu}m & & \ce{$A$} (\ce{cm\, molecule^{-1}})  & Correction Factor & References
\\     
\hline 

2.70 & \ce{$A$_{12CO2}} & 2.1 $\times$ \ce{10^{-18}} & 1.4 &  \citet{gerakines1995} (corrected) \\

4.27 & \ce{$A$_{12CO2}} & 1.1 $\times$ \ce{10^{-16}} & 1.45 & \citet{gerakines1995} (corrected) \\

15.2 & \ce{$A$_{12CO2}} & 1.6 $\times$ \ce{10^{-17}}  & 1.45 & \citet{gerakines1995} (corrected) \\

4.39 & \ce{$A$_{13CO2}} & 1.15 $\times$ \ce{10^{-16}}  & 1.47 & \citet{gerakines1995} (corrected) \\
\hline

2.35 & \ce{$A$_{12CO}} & 2.1 $\times$ \ce{10^{-19}} & 1.31 & \citet{gerakines2005} (corrected) \\

4.67 & \ce{$A$_{12CO}} & 1.4 $\times$ \ce{10^{-17}} & 1.27 &  \citet{gerakines1995} (corrected) \\

4.78 & \ce{$A$_{13CO}} & 1.7 $\times$ \ce{10^{-17}}  & 1.31 & \citet{gerakines1995} (corrected) \\

\hline
\end{tabular}
\label{Tab:bandstrengths1}
\begin{tablenotes}\footnotesize
\item{\textbf{Notes.} The corrected values of \citet{gerakines1995, gerakines2005} were taken from \citet{bouilloud2015}. The column densities derived in this work can be multiplied by the correction factors to compare with values derived using the band strengths reported in \citet{gerakines1995} and \citet{gerakines1995}. } 
\end{tablenotes}
\end{table*}  
\end{center}

We determined column densities and derived the carbon isotope ratio for the \ce{^{12}CO_2} \ce{\nu_1} + \ce{\nu_2} combination mode (2.70 \ce{\mu}m), the \ce{^{12}CO_2} \ce{\nu_3} stretching mode (4.27 \ce{\mu}m), the \ce{^{13}CO_2} \ce{\nu_3} stretching mode (4.39 \ce{\mu}m), the \ce{^{12}CO_2}  \ce{\nu_2} bending mode (15.2 \ce{\mu}m), the \ce{^{12}CO} 2-0 overtone mode (2.35 \ce{\mu}m), the \ce{^{12}CO} 1-0 stretching mode (4.67 \ce{\mu}m) and the \ce{^{13}CO} 1-0 stretching mode (4.78 \ce{\mu}m). We also detect the \ce{^{12}CO_2} 2\ce{\nu_1} + \ce{\nu_2} combination mode (2.77 \ce{\mu}m) in a number of sources. The list of sources, their luminosities, distances and the column densities for each individual vibrational mode are presented in Tables \ref{Tab:CD-CO2} and \ref{Tab:CD-CO} for \ce{CO_2} and CO, respectively.

\hfill

\subsection{\ce{CO_2}}

The 2.70 \ce{\mu}m, 4.27 \ce{\mu}m and 15.2 \ce{\mu}m vibrational modes of \ce{^{12}CO_2} are collectively detected in a number of sources. For these sources we derive \ce{^{12}CO_2}/\ce{^{13}CO_2} ratios and find values that are in agreement within the reported errorbars (Table  \ref{Tab:CD-CO2}). The largest source of uncertainty in the derived ratios stems from the error on the band strengths, the observational errors in contrast are negligible ($\sim$ 3\%). For the error on the band strengths we account for the change in the band strength due to the temperature and due to the ice mixture, more specifically a water-rich ice mixture since it is known that $\sim$ 50\% of the \ce{CO_2} resides in a water-rich matrix \citep{pontoppidan2003m,brunken2024}. 

The \ce{CO_2}-\ce{H_2O} binary ices affect each vibrational mode of \ce{CO_2} differently. The band strengths of the 2.70 \ce{\mu}m, 4.27 \ce{\mu}m and 4.39 \ce{\mu}m bands for instance decrease relative to the band strength of pure \ce{CO_2} when \ce{CO_2} is mixed with water. This subsequently result in higher \ce{^{12}CO_2} and \ce{^{13}CO_2} column densities \citep{gerakines1995}. The band strength of the 15.2 \ce{\mu}m band in contrast increases by a significant amount relative to the band strength of pure \ce{CO_2} resulting in lower \ce{^{12}CO_2} column densities. This therefore introduces a systematic error when calculating ratios from the different vibrational modes since ratios derived from the 15.2 \ce{\mu}m band will in general be lower than ratios derived from the other \ce{^{12}CO_2} vibrational modes if a band strength for a \ce{CO_2}-\ce{H_2O} mixture is used. In this study we use the band strength of pure \ce{CO_2} and account for this systematic error in the error analysis. For a detail analysis of these uncertainties we refer the reader to Appendix \ref{Appendix:A}.

\begin{figure*}[h!]
    \centering
     \includegraphics[width=1\hsize]{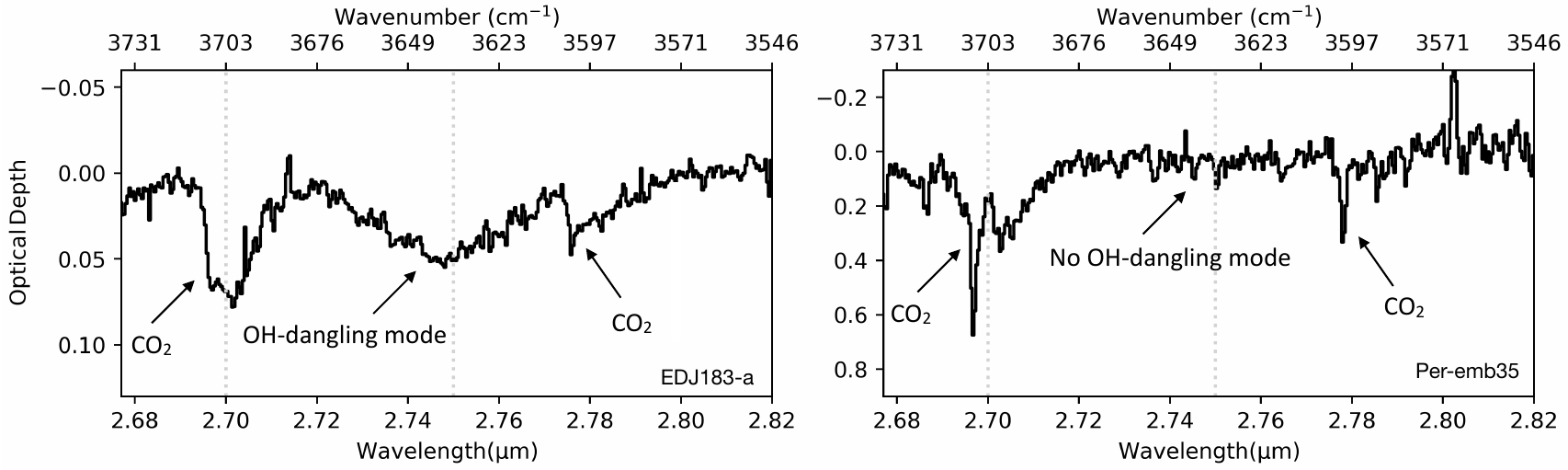}
    \caption{\ce{^{12}CO_2} combination mode (2.70 \ce{\mu}m) and water OH-dangling mode (2.75 \ce{\mu}m) for the sources EDJ183-a (not heated) and Per-emb35 (heated). The feature at 2.75 \ce{\mu}m is not detected in Per-emb35 while the 2.70 \ce{\mu}m feature is detected in both sources.}
    \label{fig:dangling-mode}
\end{figure*}

\begin{figure*}[h!]
    \centering
     \includegraphics[width=1\hsize]{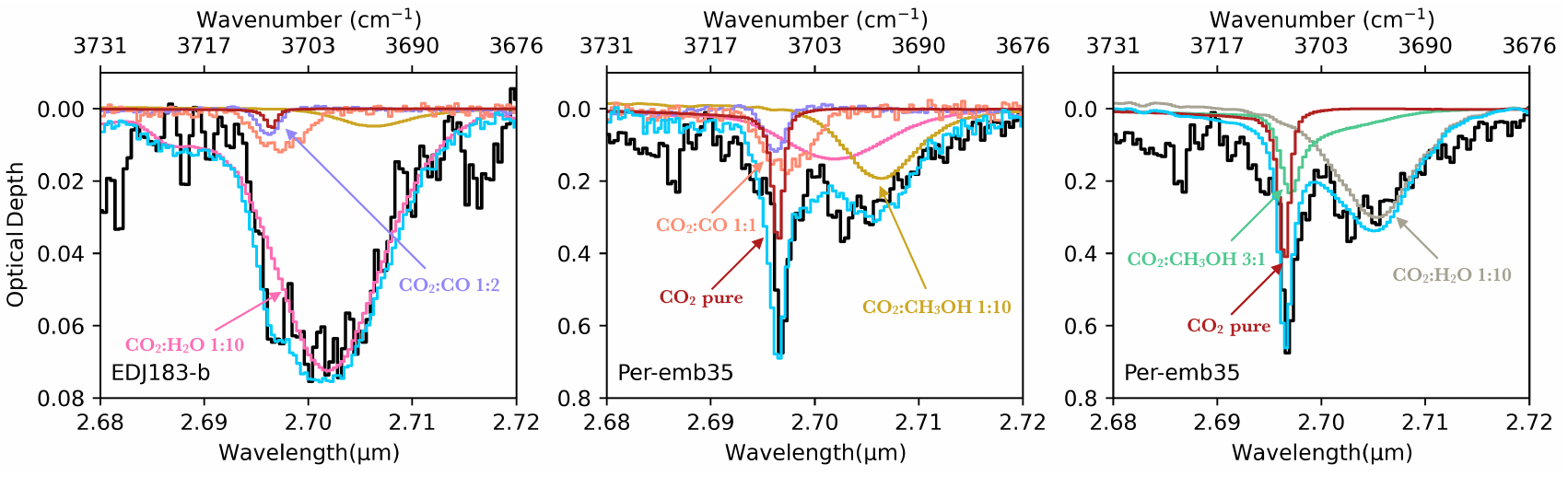}
    \caption{Profile decomposition of the \ce{^{12}CO_2} 2.70 \ce{\mu}m combination mode using selected laboratory spectra based on the analysis performed in \citet{brunken2024}. Left and middle panel: The black line shows the observed spectrum and the blue line shows the linear combination of all the five different components at low temperature. The yellow line corresponds to the \ce{CO_2}:\ce{CH_3OH} (10 K) component. The pink line shows the contribution of the broad \ce{CO_2}:\ce{H_2O} (10 K) component. The purple line corresponds to the diluted \ce{CO_2}:\ce{CO} (25 K) component while the orange line shows the contribution of \ce{CO_2} and \ce{CO} mixed in equal ratio (15 K). Finally, the dark red line corresponds to the pure \ce{CO_2} (80 K) component. Right panel: The black line shows the observed spectrum, and the blue line shows the linear combination of all the three different components at high temperatures. The green line corresponds to the hot \ce{CO_2}:\ce{CH_3OH} (105 K) component. The grey line shows the contribution of the hot \ce{CO_2}:\ce{H_2O} (160 K) component. Finally, the dark red line corresponds to the pure \ce{CO_2} (80 K) component. All laboratory spectra are taken from \citet{ehrenfreund1997infrared,ehrenfreund1999laboratory, van2006infrared} and are publicly available on the Leiden Ice Database (LIDA) \citep{rocha2022lida}. Further details on the laboratory spectra can be found in \citet{brunken2024}.}
    \label{fig:combi-fits}
\end{figure*}

\begin{figure*}[h!]
    \centering
     \includegraphics[width=0.6\hsize]{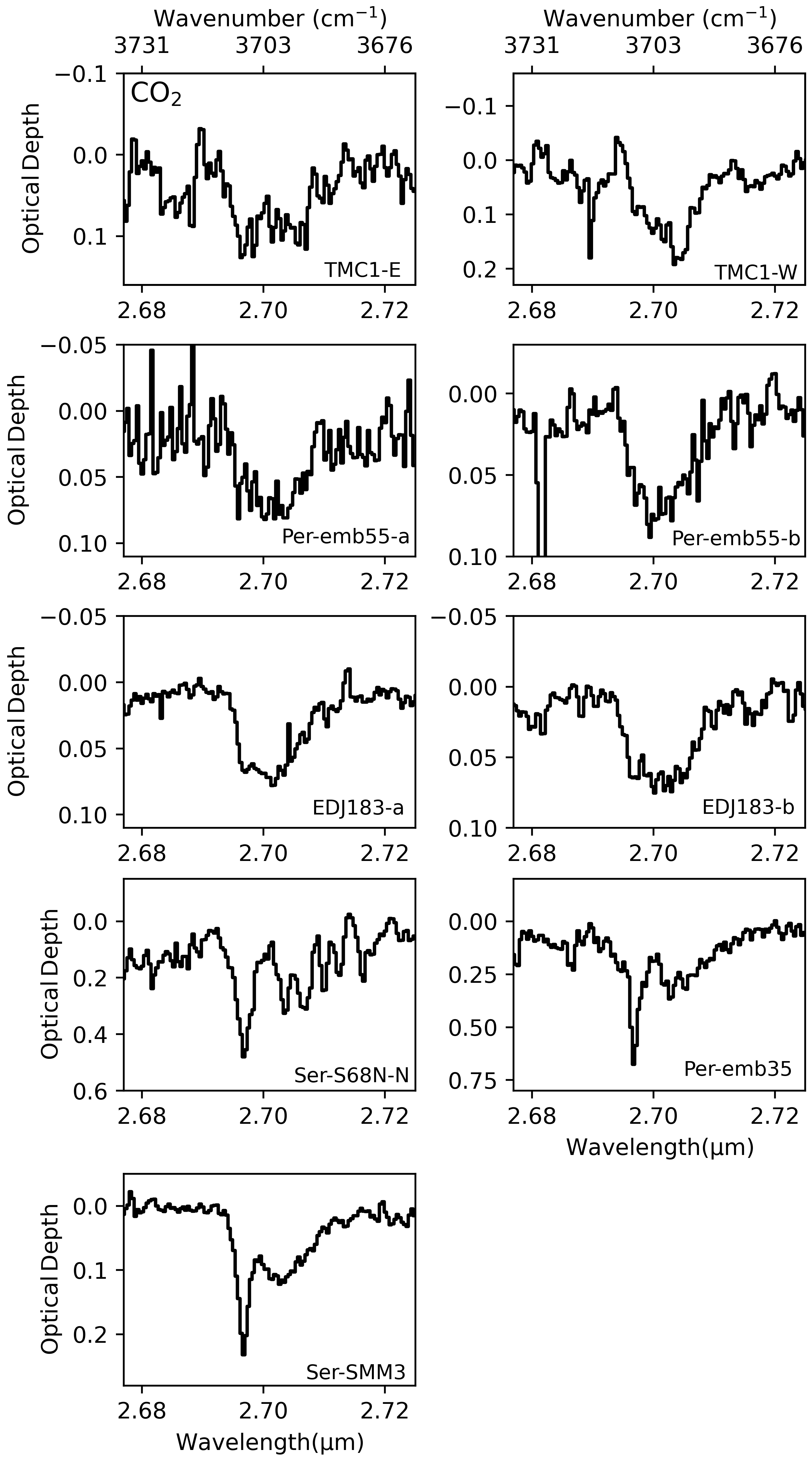}
    \caption{Overview of the \ce{^{12}CO_2} combination modes. }
    \label{fig:bandcomb}
\end{figure*}

\subsubsection{The \ce{^{12}CO_2} 2.70 \ce{\mu}m and 2.77 \ce{\mu}m combination modes}
\label{subsec:combi}

The assignment of the 2.70 \ce{\mu}m band has been the topic of multiple studies. \cite{keane2001} for instance first fitted this feature with laboratory spectra of CO-\ce{H_2O} (100:1) ices but later assigned this feature to \ce{CO_2} since the fraction of CO in the binary CO-\ce{H_2O} ices was inconsistent with the amount of CO present in the other bands. The band has also been assigned to the OH-dangling mode of porous water ice since water has multiple weak features in this spectral region \citep{noble2024}. In our sample we detect this 2.70 \ce{\mu}m feature in 9 out of 17 sources on the short-wavelength side of the OH-dangling mode situated at 2.75 \ce{\mu}m (Figure \ref{fig:dangling-mode}) and we assign this band to the combination mode of \ce{CO_2} for the following reasons. 

First, we fitted several spectra of pure \ce{CO_2} ice and \ce{CO_2} in mixtures with CO, \ce{H_2O} and \ce{CH_3OH} \citep{pontoppidan2008c2d} and we find an absorption feature at 2.70 \ce{\mu}m for every mixture containing \ce{CO_2} as illustrated in Figure \ref{fig:combi-fits}. The dangling modes of pure porous water ice and water ice mixed with other ice species do not sufficiently fit the observed profile of this feature. Additionally, we detect the OH-dangling mode of water ice interacting with other species at 2.75 \ce{\mu}m in every source where the combination mode is present except in Per-emb35 where it disappears (Figure \ref{fig:dangling-mode}). This source shows clear signs of thermal processing with double-peak profiles at 4.39 \ce{\mu}m (\ce{^{13}CO_2}) \citep{brunken2024} and 15.2 \ce{\mu}m (\ce{^{12}CO_2}) \citep{pontoppidan2008c2d} (Figures \ref{fig:cont15micron2}, \ref{fig:cont15micron3}, \ref{fig:bands4micron2} and \ref{fig:bands4micron3}).

\begin{figure*}[!ht]
    \centering
     \includegraphics[width=0.7\hsize]{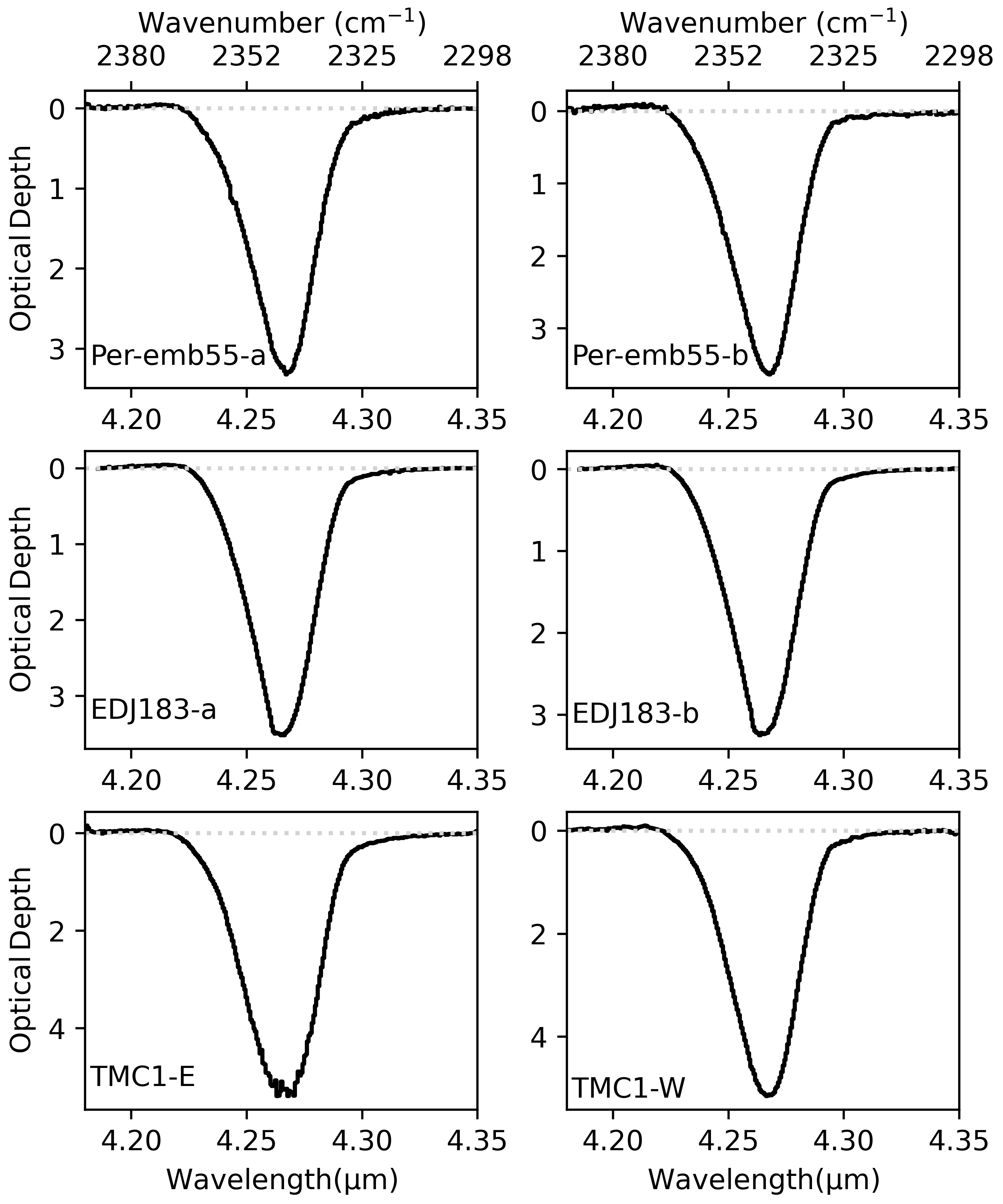}
    \caption{Overview of the \ce{^{12}CO_2} 4.27 \ce{\mu}m stretching mode. The features are shown for sources for which the band is not saturated. }
    \label{fig:co2-stretch-overview}
\end{figure*}

OH-dangling modes originate from non-H bonded OH-groups of cold amorphous water ice that is either porous or in mixtures with other species that disrupt the H-bonding network \citep{keane2001, noble2024} and experimental studies have shown that the strength of these bands decreases with increasing temperature \citep{isokoski2014}. This is because the elevated temperatures will cause the molecules to restructure and form bonds which in turn will lead to a decrease in the number of free OH-groups and the strength of the OH-dangling modes. Gradual decrease of OH trapping sites due to ice heating was also shown in experimental results by \citet{sandford1988} where the \ce{H_2O}-CO shoulder observed at 4.64 \ce{\mu}m disappeared once the amorphous \ce{H_2O} ice structure was annealed. Since the 4.39 \ce{\mu}m and 15.2 \ce{\mu}m bands of Per-emb35 show clear signs of thermal processing, the absence of the 2.75 \ce{\mu}m band in this source is therefore likely the result of the amorphous porous water structure disappearing due to the elevated temperatures in the envelope.

If both the 2.70 \ce{\mu}m and the 2.75 \ce{\mu}m bands were indeed water OH-dangling modes however, then the 2.70 \ce{\mu}m feature should also disappear when the ices undergo thermal processing. Instead what we observe is a sharpening of the profile and the appearance of a sharp peak on the short-wavelength side of the band which we successfully reproduced with the spectrum of pure \ce{CO_2} and the alternative spectral decomposition for warm ices described in \citet{brunken2024} (Figure \ref{fig:combi-fits}). This behavior is similar to the behavior observed at 4.39 \ce{\mu}m and 15.2 \ce{\mu}m when pure \ce{CO_2} ice begins to segregate from the ice matrix. This strongly indicates that the 2.70 \ce{\mu}m feature therefore cannot be primarily the product of OH-dangling modes. We do note that the 2.75 \ce{\mu}m OH-dangling mode is detected in the extended sources Ser-S68N-N and Ser-SMM3 (Figure \ref{fig:secondcombi} where the ices are also processed but this could be due to the fact that we might be tracing a fraction of cold ice components in the extended envelopes of these sources. 

Finally, while there have been laboratory spectra showing OH-dangling bonds at 2.70 \ce{\mu}m for water ice with high porosity, we note that the method of depositing gas-phase water to form very porous water is inconsistent with the mechanism of interstellar ice formation through atom addition which produces more compact ices. Therefore, in this work we assign the 2.70 \ce{\mu}m feature to the combination mode of \ce{CO_2} and quantify the column density of \ce{^{12}CO_2} from this band. The ratios extracted from this band are presented in Table \ref{Tab:CD-CO2}. 

We note that the column densities we derive from the 2.70 \ce{\mu}m band are consistent with the column densities extracted from the other two \ce{^{12}CO_2} vibrational modes in both our cold and heated sources except for Ser-SMM3. This is further evidence that the majority of this band is indeed for the most part the \ce{^{12}CO_2} combination mode.

We also observe the 2\ce{\nu_1} + \ce{\nu_2} combination mode at 2.77 \ce{\mu}m in Per-emb35 and Ser-SMM3 (Figure \ref{fig:secondcombi}). The band strength of this feature is however a factor $\sim$ 3 lower than that of the 2.70 \ce{\mu}m and it is also highly susceptible to changes in the temperature, even more so than its counterpart at 2.70 \ce{\mu}m. Because of these high uncertainties we will refrain from quantifying carbon isotope ratios from this weak feature.

\subsubsection{The \ce{^{12}CO_2} 4.27 \ce{\mu}m stretching mode}

In Figure \ref{fig:co2-stretch-overview} we present the 4.27 \ce{\mu}m stretching mode for 6 out of 17 sources. While this band is detected in the remaining sources, they all suffer from heavy saturation ($\tau$ > 5 ). As a test for using the 4.27 \ce{\mu}m band for saturated sources, we fitted the laboratory spectrum of \ce{CO_2}:\ce{H_2O} 1:10 at 10 K \citep{ehrenfreund1999laboratory} to the wings of the 4.27 \ce{\mu}m band in Per-emb35 using the 15.2 \ce{\mu}m bending mode as an anchor.

Overall the band profiles of the selected 6 sources appear to be very similar with only the slope of the short-wavelength wing varying depending on the source. Additionally, the strength of the scattering feature on the short-wavelength wing differs per source. The column densities derived from this band are in good agreement with the column densities derived from the 2.70 \ce{\mu}m combination mode for each source. 

For the fitted band of Per-emb35 we find a \ce{^{12}CO_2}/\ce{^{13}CO_2} ratio of $\sim$ 80 $\pm$ 21 (Figure \ref{fig:per35-str}). This value is consistent within the reported errorbars with the ratios derived from the other two vibrational modes in this same source. Due to the ambiguity of the true band profiles and optical depths of these saturated bands however, we will omit the band of these sources from this study.

\subsubsection{The \ce{^{12}CO_2} 15.2 \ce{\mu}m bending mode}
\label{subsubsec:bending}

The 15.2 \ce{\mu}m bending mode is detected in all 17 sources, each showing distinctive band profiles (Figures \ref{fig:cont15micron1}, \ref{fig:cont15micron2},  \ref{fig:cont15micron3} and \ref{fig:cont15micron4}). Most notable are the sources that have the double-peak profile, a telltale sign of ice heating \citep{ehrenfreund1998ice,pontoppidan2008c2d}. 

In Section \ref{sec:cont-15} we discussed the continuum subtraction method and the extended wing-profile after subtracting the silicate model. We investigated the effect of this extended wing in comparison to the wing profile of \citet{pontoppidan2008c2d} that extends to $\sim$ 16 \ce{\mu}m instead of $\sim$ 18 \ce{\mu}m, and found a difference in column density of $\sim$ 9\%, which is negligible compared to the uncertainties of the band strengths (see also Appendix \ref{Appendix:A}). The column densities derived from the bending mode are presented in Table \ref{Tab:CD-CO2}.

\subsubsection{The \ce{^{13}CO_2} 4.39 \ce{\mu}m stretching mode}

\ce{^{13}CO_2} column densities were derived from the 4.39 \ce{\mu}m band for all 17 sources (Table \ref{Tab:CD-CO2}). Consistent with the 15.2 \ce{\mu}m band, we observe a double-peak profile at 4.39 \ce{\mu}m for every source that displays peak splitting at 15.2 \ce{\mu}m. One hindrance is the CO line forest that bleeds into the \ce{^{13}CO_2} band of several sources in this sample, potentially causing a slight underestimation of the \ce{^{13}CO_2} column densities. We briefly investigated the effect of the gaseous CO lines in EDJ183-b by fitting a Gaussian profile to the bottom of the \ce{^{13}CO_2} band and quantifying the column density from this curve. We find that the \ce{^{12}CO_2}/\ce{^{13}CO_2} ratio decreases by 35\% if the integrated optical depth is determined from the Gaussian curve instead of the band directly. In Figure \ref{fig:bands13CO2} we show the Gaussian curves that were ultimately used to determine the integrated optical depths for the \ce{^{13}CO_2} bands most affected by the gaseous CO lines. For the sources where the \ce{^{13}CO_2} features were well isolated from the line forest the integrated optical depths were determined from the ice bands directly.

\begin{figure}[h!]
    \centering
     \includegraphics[width=0.8\hsize]{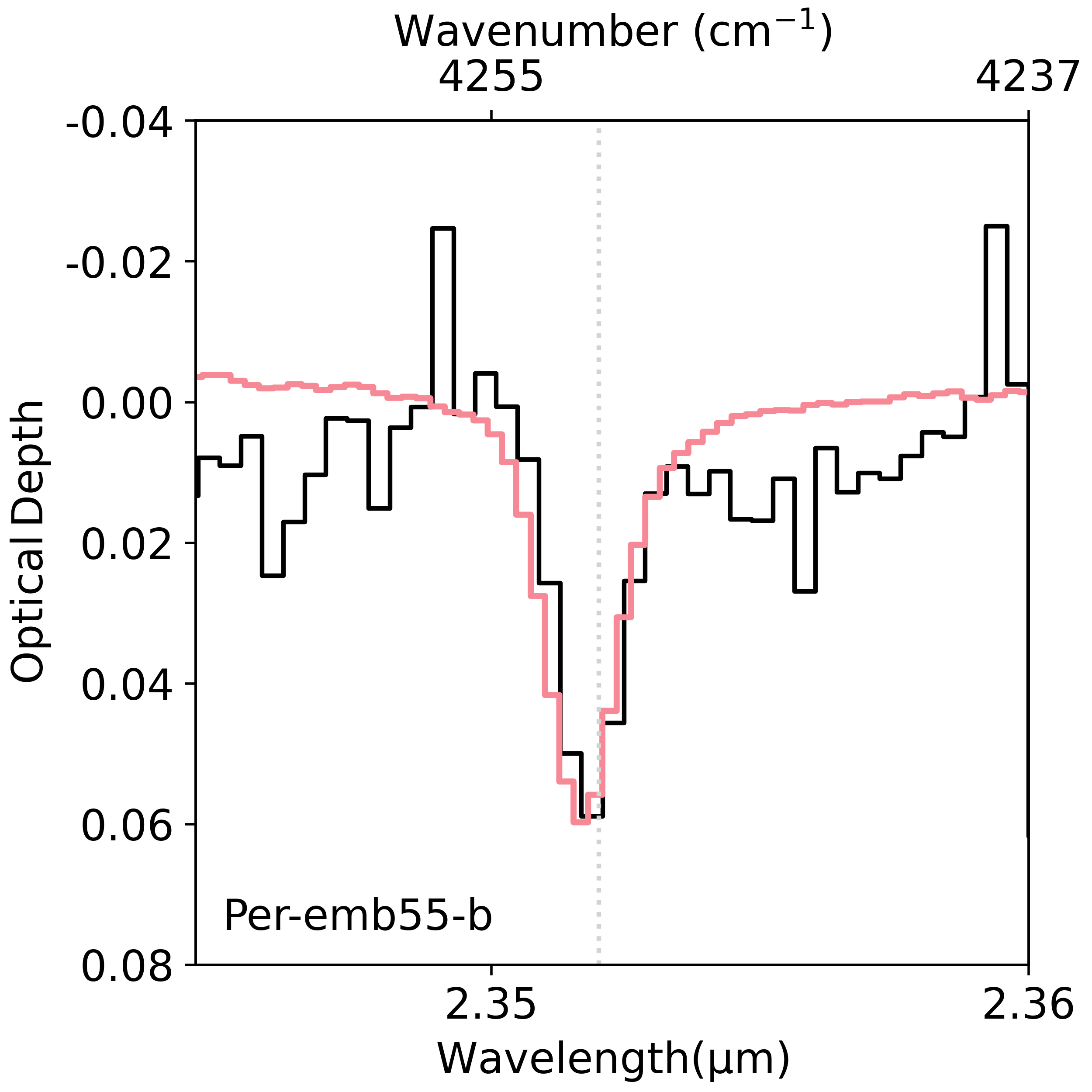}
    \caption{The 2.35 \ce{\mu}m feature detected in the source Per-emb55-b. The laboratory spectrum of pure CO is plotted over the \ce{\nu} = 2-0 absorption feature (pink). }
    \label{fig:overtone-per}
\end{figure}

\begin{center}
\begin{table*}[hbt!]

\caption{Column densities and isotope ratios derived for \ce{^{12}CO_2} and \ce{^{13}CO_2} ice.}
\small
\centering
\scalebox{0.8}{\begin{tabular}{lccccccccc}
\hline \hline
Source & L ($L_{\odot}$) & Distance (pc)  & N \ce{^{12}CO_2} 2.70 \ce{\mu}m & N \ce{^{12}CO_2} 4.27 \ce{\mu}m &	N \ce{^{13}CO_2} 4.39 \ce{\mu}m  & N \ce{^{12}CO_2} 15.2 \ce{\mu}m  &	\ce{^{12}C}/\ce{^{13}C} 2.70 \ce{\mu}m &	\ce{^{12}C}/\ce{^{13}C} 4.27 \ce{\mu}m & \ce{^{12}C}/\ce{^{13}C} 15.2 \ce{\mu}m 
\\     
\hline

B1-b	& 0.2 & 293 & -	& - & 2.6 $\times$ \ce{10^{16}} & 2.6 $\times$ \ce{10^{18}}  & - &-& 100 $\pm$ 46 \\

TMC1-E & 0.7 & 142 & 1.0  $\times$ \ce{10^{18}} & 1.1 $\times$ \ce{10^{18}} &	1.6 $\times$ \ce{10^{16}} & 	1.2 $\times$ \ce{10^{18}}  &	62 $\pm$ 19 & 69 $\pm$ 18 &	75 $\pm$ 34 \\

TMC1-W & 0.7 & 142 & 1.3 $\times$ \ce{10^{18}} & 9.0 $\times$ \ce{10^{17}} &	1.2 $\times$ \ce{10^{16}} &	1.2 $\times$ \ce{10^{18}}  &	81 $\pm$ 25 & 58 $\pm$ 15 &	76 $\pm$ 35 \\

B1-a-N & 1.5 & 293 & - & - & 2.2$\times$ \ce{10^{16}} & 	2.1 $\times$ \ce{10^{18}} &	- &- & 94 $\pm$ 43 \\

B1-a-S & 1.5 & 293 &	- & - & 1.9 $\times$ \ce{10^{16}} &	2.1 $\times$ \ce{10^{18}}  &	- &-&	114 $\pm$ 52 \\

Per-emb55-a & 1.8 & 293 &	5.7 $\times$ \ce{10^{17}} & 5.7 \ce{10^{17}} &	8.1 $\times$ \ce{10^{15}} &	6.0 $\times$ \ce{10^{17}}  &	70 $\pm$ 21 & 70 $\pm$ 18 &	74 $\pm$ 34 \\

Per-emb55-b & 1.8 & 293 &	5.8 $\times$ \ce{10^{17}} & 6.2 \ce{10^{17}}
 &	7.7 $\times$ \ce{10^{15}} &	5.5 $\times$ \ce{10^{17}} &	74 $\pm$ 23  & 80 $\pm$ 21 &	70 $\pm$ 32 \\

L1527 & 3.1 & 142 & - & - &	4.6 $\times$ \ce{10^{16}} &	4.8 $\times$ \ce{10^{18}}  &	- & -&	105 $\pm$ 48 \\

B1-c	& 3.2 & 293 &  - & - & 7.1 $\times$ \ce{10^{16}} & 7.9 $\times$ \ce{10^{18}}  & -  &-& 112 $\pm$ 52 \\

EDJ183-a & 3.2 & 293  & 5.7 $\times$ \ce{10^{17}} & 5.9 $\times$ \ce{10^{17}} &	7.0 $\times$ \ce{10^{15}} &	5.2 $\times$ \ce{10^{17}}	& 81 $\pm$ 25 & 84 $\pm$ 22 &	74 $\pm$ 34\\

EDJ183-b & 3.2 & 293 & 5.8 $\times$ \ce{10^{17}} & 5.5 $\times$ \ce{10^{17}}
 & 5.8 $\times$ \ce{10^{15}}	 & 5.1 $\times$ \ce{10^{17}}  &	99 $\pm$ 31 & 94 $\pm$ 25 & 88 $\pm$ 40\\

Per-emb22 & 3.6 & 293 & - &	 - &4.4 $\times$ \ce{10^{16}}  &	5.6 $\times$ \ce{10^{18}}  &	- &-&	125 $\pm$ 58 \\

Ser-S68N-N  &6 & 435 &	2.8 $\times$ \ce{10^{18}} & - &	2.6 $\times$ \ce{10^{16}} &	3.1 $\times$ \ce{10^{18}} &	 108 $\pm$ 33 &-& 120 $\pm$ 55  \\

Per-emb33 & 8.3 & 293 &	- & - &	6.1 $\times$ \ce{10^{16}}&	6.4  $\times$ \ce{10^{18}}  &	- &-&	105 $\pm$ 48\\

Per-emb35 & 9.1 & 293 &	3.7  $\times$ \ce{10^{18}} & - &	2.8  $\times$ \ce{10^{16}} &	2.8  $\times$ \ce{10^{18}}  &	132 $\pm$ 41 &-&	99 $\pm$ 48 \\

Ser-SMM3 & 28 & 435 &1.1 $\times$ \ce{10^{18}} & - & 2.0 $\times$ \ce{10^{16}} &	2.2  $\times$ \ce{10^{18}}  &	55 $\pm$ 17 & -&	108 $\pm$ 50\\

Per-emb27 & 36 & 293 &	- & - & 5.9 $\times$ \ce{10^{16}} &	6.5  $\times$ \ce{10^{18}}  &	- & -&	110 $\pm$ 51\\

\hline
\end{tabular}}
\label{Tab:CD-CO2}
\begin{tablenotes}\footnotesize
\item{\textbf{Notes.}The main source of uncertainty in the derived ratios stem from the error on the band strength, the observational errors in contrast are negligible ($\sim$ 3\%). The error on the band strength include the change in the band strength due to temperature and due to a water-rich ice mixture. The error analysis is presented in detail in Appendix \ref{Appendix:A}.} 
\end{tablenotes}

\end{table*} 

\end{center}

\begin{figure*}[!ht]
    \centering
     \includegraphics[width=0.9\hsize]{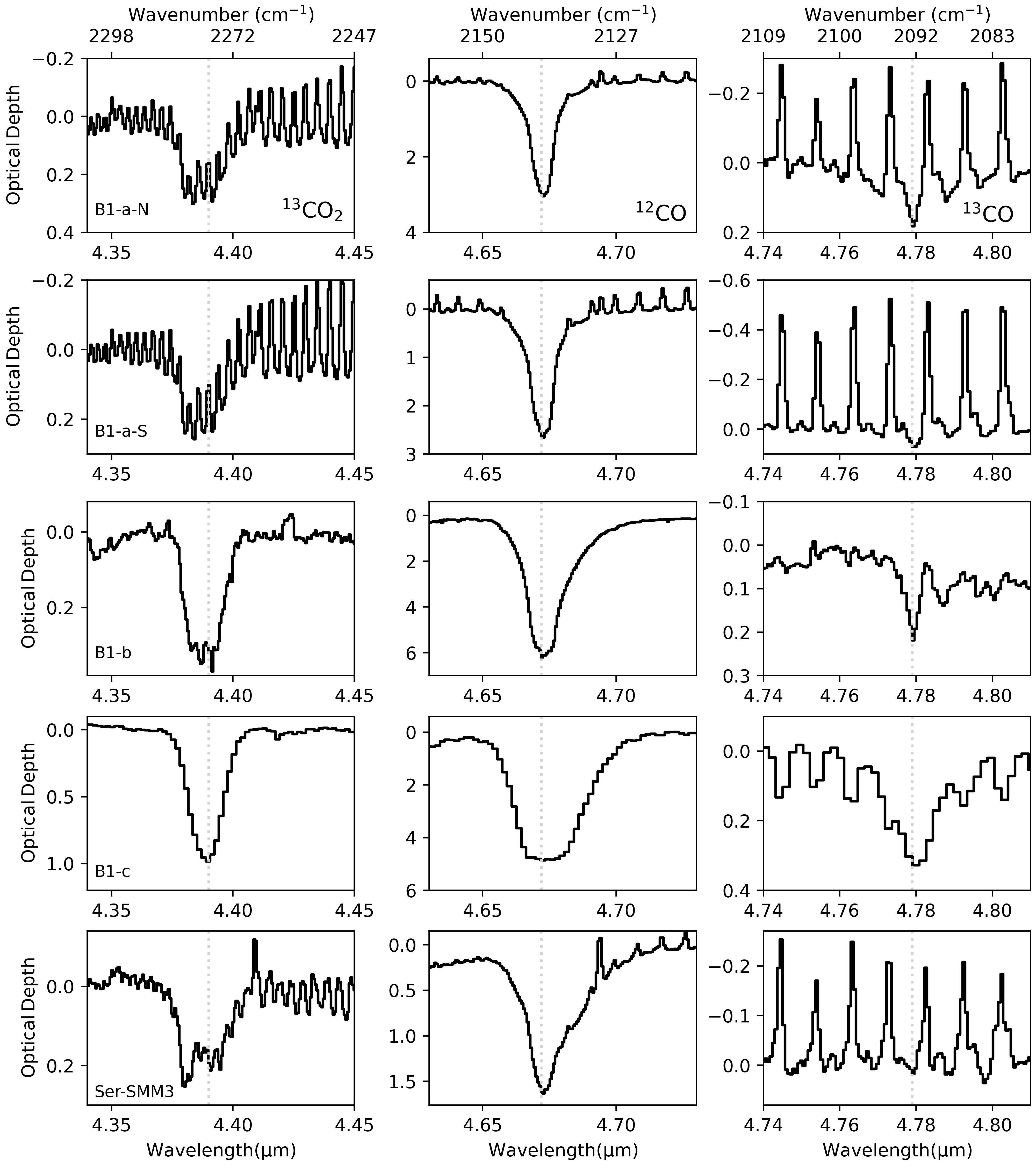}
    \caption{Overview of the \ce{^{13}CO_2} (left), \ce{^{12}CO} (middle) and \ce{^{13}CO} (right) stretching modes in the 4 \ce{\mu}m region. The \ce{^{12}CO} gas phase lines are seen superimposed in several sources. }
    \label{fig:bands4micron1}
\end{figure*}

\begin{figure*}[!ht]
    \centering
     \includegraphics[width=0.7\hsize]{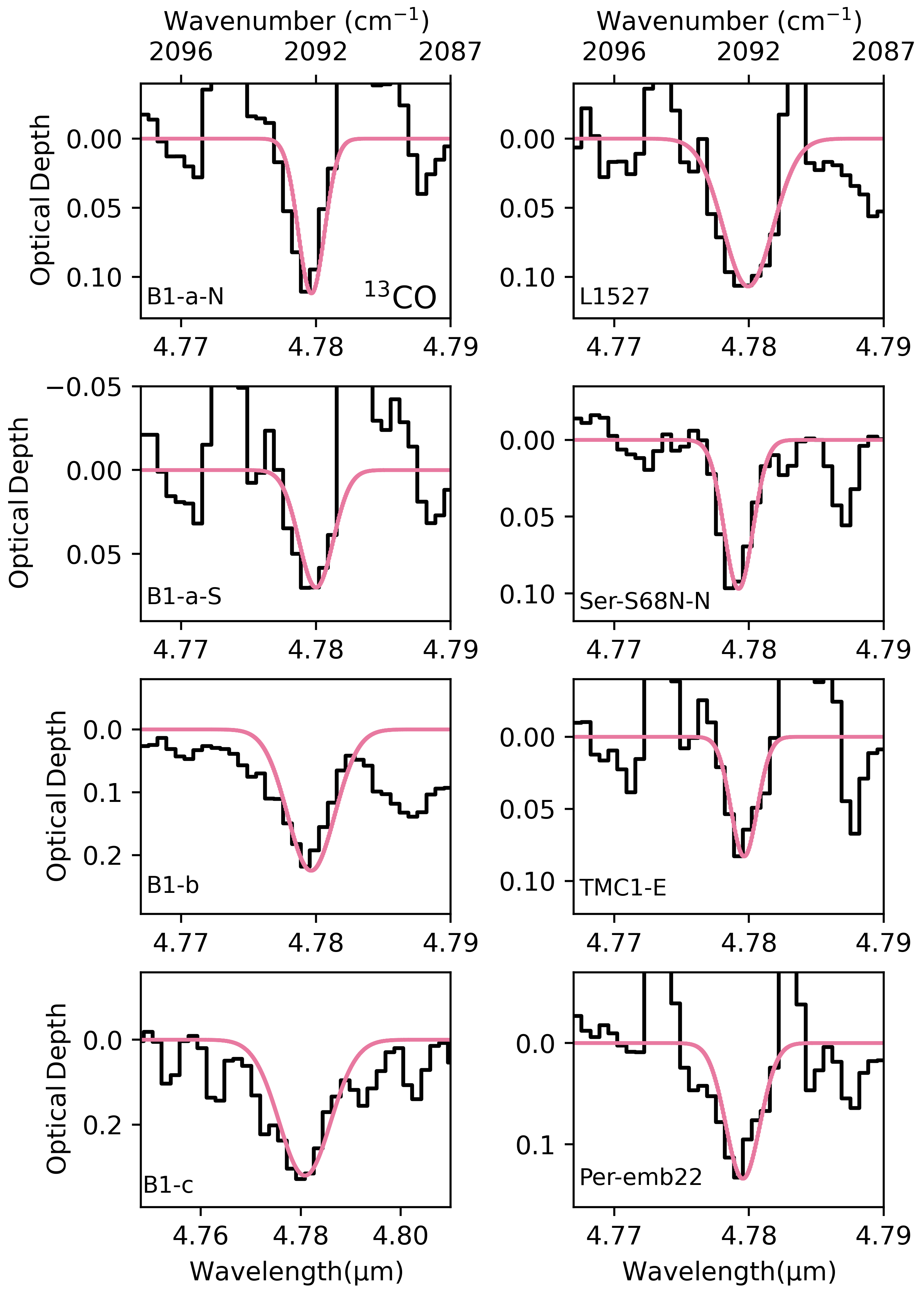}
    \caption{A zoom in of the \ce{^{13}CO} absorption features. The Gaussian curves (pink) fitted to the bands illustrate the bounds used to determine the integrated optical depths. }
    \label{fig:bands13CO}
\end{figure*}

\subsubsection{\ce{^{12}CO_2}/\ce{^{13}CO_2}}

In Table \ref{Tab:CD-CO2} we compare the column densities and ratios derived from the different \ce{CO_2} ice features. For the sources in which the 2.70 \ce{\mu}m combination mode, the 4.27 \ce{\mu}m stretching mode and the 15.2 \ce{\mu}m bending mode are detected, we find consistent ratios except for TMC1-W where the ratio derived from the 4.27 \ce{\mu}m stretching mode is $\sim$ 25\% lower than the values extracted from both the 2.70 \ce{\mu}m and the 15.2 \ce{\mu}m bands. This deviation still lies within our reported errorbars however. We also find a discrepancy of $\sim$ 50\% between the ratio derived from the 2.70 \ce{\mu}m and the 15.2 \ce{\mu}m bands in Ser-SMM3. 

For the 2.70 \ce{\mu}m band, the 4.27 \ce{\mu}m band and the 15.2 \ce{\mu}m band we find mean \ce{^{12}CO_2}/\ce{^{13}CO_2} ratios of 85 $\pm$ 23, 76 $\pm$ 12 and 97 $\pm$ 17, respectively. These are similar to the median values we derived from these vibrational modes (see also Section \ref{subsec:co2solid}). We note that the variation in ratios derived for the different sources and the uncertainty margin on the mean ratios could in fact be due to genuine differences that reflect the different chemical conditions of these systems.

\begin{center}
\begin{table*}[hbt!]
\caption{Column densities and isotope ratios derived for \ce{^{12}CO} and \ce{^{13}CO} ice.}
\small
\centering
\scalebox{0.9}{\begin{tabular}{lccccccc}
\hline \hline
Source & L ($L_{\odot}$) & Distance (pc) & N \ce{^{12}CO} 2.35 \ce{\mu}m & N \ce{^{12}CO} 4.67 \ce{\mu}m & N \ce{^{13}CO} 4.78 \ce{\mu}m &	\ce{^{12}C}/\ce{^{13}C} 2.35 \ce{\mu}m & \ce{^{12}C}/\ce{^{13}C} 4.67 \ce{\mu}m
\\     
\hline

B1-b & 0.2 & 293 &	-	& 4.1  $\times$ \ce{10^{18}}  &	2.5  $\times$ \ce{10^{16}} &	- &	162 $\pm$ 24 \\

TMC1-E & 0.7 & 142 & - & 8.1  $\times$ \ce{10^{17}} &	5.2 $\times$ \ce{10^{15}} &- &	156 $\pm$ 23 \\

TMC1-W & 0.7 & 142 & - & 	8.2 $\times$ \ce{10^{17}} &	-	& - &	- \\

B1-a-N	& 1.5 & 293 & -	 &1.2  $\times$ \ce{10^{18}} &	7.0  $\times$ \ce{10^{15}}  &	- &	171 $\pm$ 26 \\

B1-a-S	& 1.5 & 293 & -&	1.1  $\times$ \ce{10^{18}} &	5.2  $\times$ \ce{10^{15}}  &	- &	205 $\pm$ 31 \\

Per-emb55-a & 1.8 & 293 & - &1.2  $\times$ \ce{10^{18}} &	6.8 $\times$ \ce{10^{15}}  &	- &	177 $\pm$ 27 \\

Per-emb55-b & 1.8 & 293 &	7.9 $\times$ \ce{10^{17}}  & 1.3 $\times$ \ce{10^{18}} &	8.2 $\times$ \ce{10^{15}}  &	96 $\pm$ 14 &	152 $\pm$ 23\\

L1527 & 3.1 & 142 & -	&2.2  $\times$ \ce{10^{18}} &	1.3 $\times$ \ce{10^{16}}  &	- &	171 $\pm$ 26 \\

B1-c	& 3.2 & 293 &  - &	5.0  $\times$ \ce{10^{18}} &	1.1  $\times$ \ce{10^{17}}	& - & 47 $\pm$ 7 \\

EDJ183-a & 3.2 & 293 & - & 	1.0 $\times$ \ce{10^{18}} &	6.8 $\times$ \ce{10^{15}} & - &	147 $\pm$ 22 \\

EDJ183-b & 3.2 & 293 &	- &	9.4 $\times$ \ce{10^{17}} &	6.2 $\times$ \ce{10^{15}} & - &	151 $\pm$ 23 \\

Per-emb22 & 3.6 & 293 & -	& 3.2  $\times$ \ce{10^{18}} &	1.1 $\times$ \ce{10^{16}}   &	- &	294 $\pm$ 44 \\

Ser-S68N-N & 6 & 435 &- & 	1.9  $\times$ \ce{10^{18}} &	6.7 $\times$ \ce{10^{15}} & - & 278 $\pm$ 42 \\

Per-emb33 &	8.3 & 293 & -	& 3.6  $\times$ \ce{10^{18}} & 8.0 $\times$ \ce{10^{15}}  & - &	447 $\pm$ 67 \\

Per-emb35 & 9.1 & 293 &	-	& 9.5  $\times$ \ce{10^{18}} & - &	- &	-  \\

Ser-SMM3 & 28 & 435  & - & 1.2  $\times$ \ce{10^{18}} & - &	-  &	- \\

Per-emb27 & 36 & 293 & 	-	& 3.4  $\times$ \ce{10^{18}} & 1.1 $\times$ \ce{10^{16}}  &- &	306 $\pm$ 46 \\

\hline
\end{tabular}}
\label{Tab:CD-CO}
\begin{tablenotes}\footnotesize
\item{\textbf{Notes.}The main source of uncertainty in the derived ratios stem from the error on the band strength, the observational errors in contrast are negligible ($\sim$ 3\%). The error on the band strength include the change in the band strength due to temperature and due to a water-rich ice mixture. The error analysis is presented in detail in Appendix \ref{Appendix:A}.} 
\end{tablenotes}
\end{table*}  
\end{center}

\subsection{CO}
\subsubsection{The \ce{^{12}CO} 2.35 \ce{\mu}m overtone mode}

We detect a weak absorption feature at 2.35 \ce{\mu}m above 3$\ce{\sigma}$ level for 6 out of the 17 sources (TMC1-E, TMC1-W, Per-emb55-b, EJD183-a, EDJ183-b and Ser-S68N-N). In previous experimental work, this band was assigned to the \ce{\nu} = 2-0 overtone mode of CO that peaks at this wavelength in the near-infrared \citep{fink1982,gerakines2005}. We used experimental data of pure CO to analyze the band profiles of these absorption features and found that the peak position of all sources except for Per-emb55-b were significantly red-shifted (Figure \ref{fig:overtone-per}).  

The sources in which the the 2.35 \ce{\mu}m features are red-shifted also display strong gas-phase CO overtone photospheric absorption lines that peak in the 2 \ce{\mu}m spectral region. These gaseous CO overtone modes originate from the central protostellar embryo and appear in absorption due to the strong thermal gradient of the gas surrounding the protostar \citep{legouellec2024}. Among these gas-phase absorption lines is the \ce{\nu} = 4-2 at 2.3525 \ce{\mu}m that overlaps with the 2.35 \ce{\mu}m CO ice absorption band causing the `red-shifted' ice bands we are observing in most of these sources.

In order to disentangle and subtract the gaseous CO overtone modes, the 2 \ce{\mu}m hot dust emission was first modelled with a blackbody emission. The photosphere was modelled using the BT-Settl grid of photospheric models  \citep{allard2012}. The Starfish modeling framework from \citet{czekala2015} was used (see also Figure \ref{fig:tmc1-a-model}). Further details on these modeling procedures are presented in Le Gouellec et al. (in prep.). The findings show that the models converged well in TMC1-E, TMC1-W, EDJ183-a, EDJ183-b and Ser-S68N-N and after subtraction no residual ice features were observed in these sources. A faint photosphere is detected in Per-emb55-a and a weak absorption feature is visible at 2.35 \ce{\mu}m. This feature is however faint and hard to separate from the noise spikes in the spectrum. In Per-emb55-b no photosphere was detected and we can conclude that the 2.35 \ce{\mu}m feature is indeed the CO overtone ice absorption band. 

Because of the gas contamination, we discarded the other five sources where these gas signatures are observed and extracted the \ce{^{12}CO} column density from the band observed in Per-emb55-b only (Figure \ref{fig:overtone-per}). The column density we derive from this band is a factor $\sim$ 1.6 lower than the column density derived from the 4.67 \ce{\mu}m band in this same source. One possible explanation for the discrepancy between these two vibrational modes is the band strength of this relatively weak absorption feature.

\subsubsection{The \ce{^{12}CO} 4.67 \ce{\mu}m stretching mode}
\label{subsec:stretching}

The bands of the \ce{^{12}CO} 4.67 \ce{\mu}m symmetric stretching mode are presented in Figures \ref{fig:bands4micron1}, \ref{fig:bands4micron2}, \ref{fig:bands4micron3} and \ref{fig:bands4micron4} and the column densities in Table \ref{Tab:CD-CO}. We find variations of more than an order of magnitude between the different sources and ascribe this to the volatility of CO. Because of its lower desorption temperature, CO ice is highly sensitive to the temperature structure of the protostellar envelope which in turn can significantly impact the CO ice budget and cause variations depending on the source.

\subsubsection{The \ce{^{13}CO} 4.78 \ce{\mu}m stretching mode}

The \ce{^{13}CO} 4.78 \ce{\mu}m stretching mode is observed in 14 out of 17 sources of which only two are well isolated from the CO gas-phase rotation-vibrational lines (B1-b and B1-c). The strong line forest that dominates most of the sources in this sample poses a challenge to the study of this weak feature since the actual shape and optical depth of this band are likely affected by the gas lines. We briefly investigated the effect of the emission lines by fitting Gaussian profiles and subtracting the two gaseous CO lines closest to the bands and we find that the \ce{^{13}CO} column density increases by a factor 1.4 compared to the non-subtracted band which is still within our reported uncertainties. We note however that the gas-subtracted \ce{^{13}CO} ice bands are significantly broader than non-subtracted bands and also much broader than the band of the laboratory spectrum of pure CO which could be the main contributor of this band \citep{boogert2002b, pontoppidan2003m}. On the other hand, a broad band profile is possible if the \ce{^{13}CO} ice is residing in a polar \ce{CH_3OH}-rich ice matrix \citep{peantado2015}. A careful modelling and removal of the gaseous CO lines is needed in order to properly isolate the ice feature and quantify the \ce{^{13}CO} ice column density. Such CO gas modeling is however outside the scope of this paper. In Figures \ref{fig:bands13CO} and \ref{fig:bands13CO-2} we show the Gaussian curves that were fitted to the \ce{^{13}CO} features in order to determine the integrated optical depths.

\subsubsection{\ce{^{12}CO}/\ce{^{13}CO}}
\label{subsub:ratio-CO}

The carbon isotope ratios derived from the CO ice absorption bands are presented in Table \ref{Tab:CD-CO}. In general we find high \ce{^{12}CO}/\ce{^{13}CO} ratios. The sources B1-a-S, Per-emb22, Per-emb33, Ser-S68N-N and Per-emb27 in particular have very high ratios, all of which also have very weak \ce{^{13}CO} ice absorption features. Moreover, the \ce{^{13}CO} ice bands of some of these sources are buried in the gas line forest which could be the reason behind some of these extreme ratios. We derive a sub-ISM ratio for the source B1-c but we note that the spectrum of this source was taken at medium resolution unlike the other sources in this sample. Consequently, the narrow \ce{^{13}CO} band might not be fully resolved. Additionally, the gaseous CO lines in this source are in absorption and could be blended with the unresolved ice feature further contributing to its width.

Finally, we derive a \ce{^{12}CO}/\ce{^{13}C} ratio of 96 $\pm$ 14 from 2.35 \ce{\mu}m overtone mode in Per-emb55-a, which is a factor $\sim$ 1.6 lower than the ratio of 152 $\pm$ 23 derived in this same source from the 4.67 \ce{\mu}m band. For the 4.67 \ce{\mu}m band we find a mean \ce{^{12}CO}/\ce{^{13}CO} ratio of 165 $\pm$ 52 for the sample as whole when we exclude the two extreme outliers Per-emb22, Per-emb33 and Per-emb27. Due to the fluctuating values derived from the CO vibrational modes, we advise the reader to refer to the ratios derived from the \ce{CO_2} vibrational modes rather than the CO ice bands.

\subsection{Limitations and future work}

As shown in the previous sections, we examined the carbon isotope ratios for multiple CO and \ce{CO_2} vibrational modes for a large sample of low mass sources. We find, for the most part, consistent values especially between the ratios derived from the \ce{CO_2} ice bands. One source of uncertainty are the band strengths especially for the weaker \ce{^{12}CO} overtone mode. The 4.40 - 4.5 \ce{\mu}m spectral region is also heavily dominated by gaseous CO rotation-vibrational lines which poses a challenge when quantifying \ce{^{13}CO} column densities from the absorption feature at 4.78 \ce{\mu}m. Therefore, a careful modeling and subtraction of these gas lines is needed to properly isolate in the \ce{^{13}CO} ice features and determine \ce{^{13}CO} column densities. Alternatively, spectral extractions at locations where the gas-phase lines are weak or absent could also aid in the analysis of the \ce{^{13}CO} ice features. Finally, future work should also focus on finding similarities or variations between close binaries as well as perform spatial ice mapping, a feat that is now possible with the JWST.

\section{Discussion}
\label{sec:4}

The carbon isotope ratio is a crucial link between the evolutionary phases of star and planet formation because of its sensitivity to the chemical and physical conditions that reign during each stage. In the following section we will examine how the results presented in this work relate to the trends observed across different astronomical environments both in the solid state and gas phase.

\begin{figure*}[h!]
    \centering
     \includegraphics[width=1\hsize]{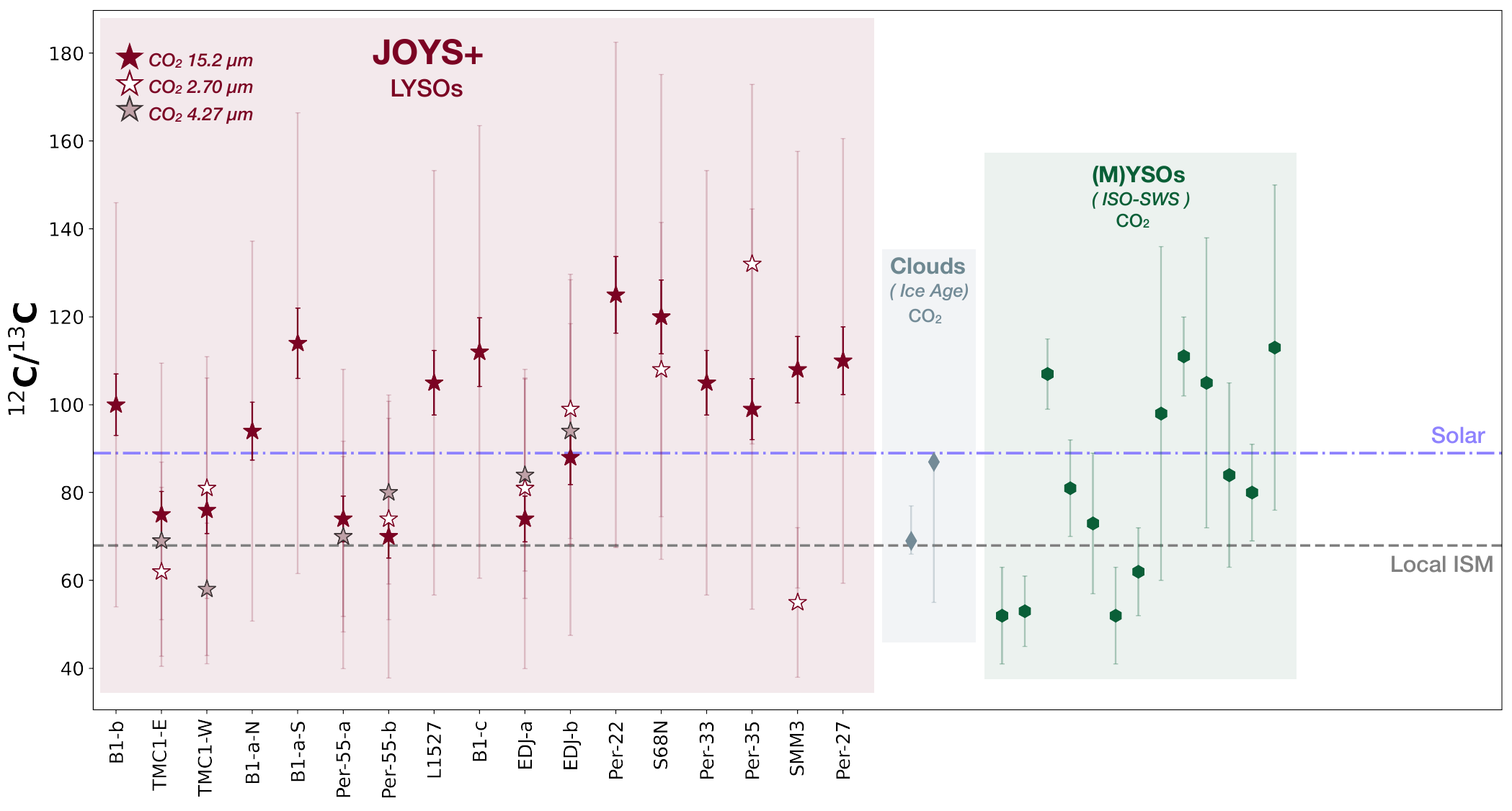}
    \caption{Solid \ce{^{12}C}/\ce{^{13}C} for various astronomical environments. Red-filled, grey-filled and white-filled stars represent ratios derived from the \ce{^{12}CO_2} 15.2\ce{\mu}m, 4.27 \ce{\mu}m and 2.70 \ce{\mu}m bands respectively in this work.  Also included are molecular clouds (Ice Age) \citep{mcclure2023ice}, massive YSO (ISO-SWS). \citep{boogert1999iso}. The dashed purple line shows the solar abundance and the dot-dashed gray line shows the local ISM ratio \citep{boogert1999iso}. The small solid errorbars of the JOYS+ data points, illustrate when the band strengths are excluded from the error analysis. The faint and generally larger errorbars of the JOYS+ data points illustrate the uncertainty including the error on the band strength. Further details on the error analysis can be found in Appendix \ref{Appendix:A}.}
    \label{fig:ratios-CO2}
\end{figure*}

\begin{figure}[h!]
    \centering
     \includegraphics[width=0.9\hsize]{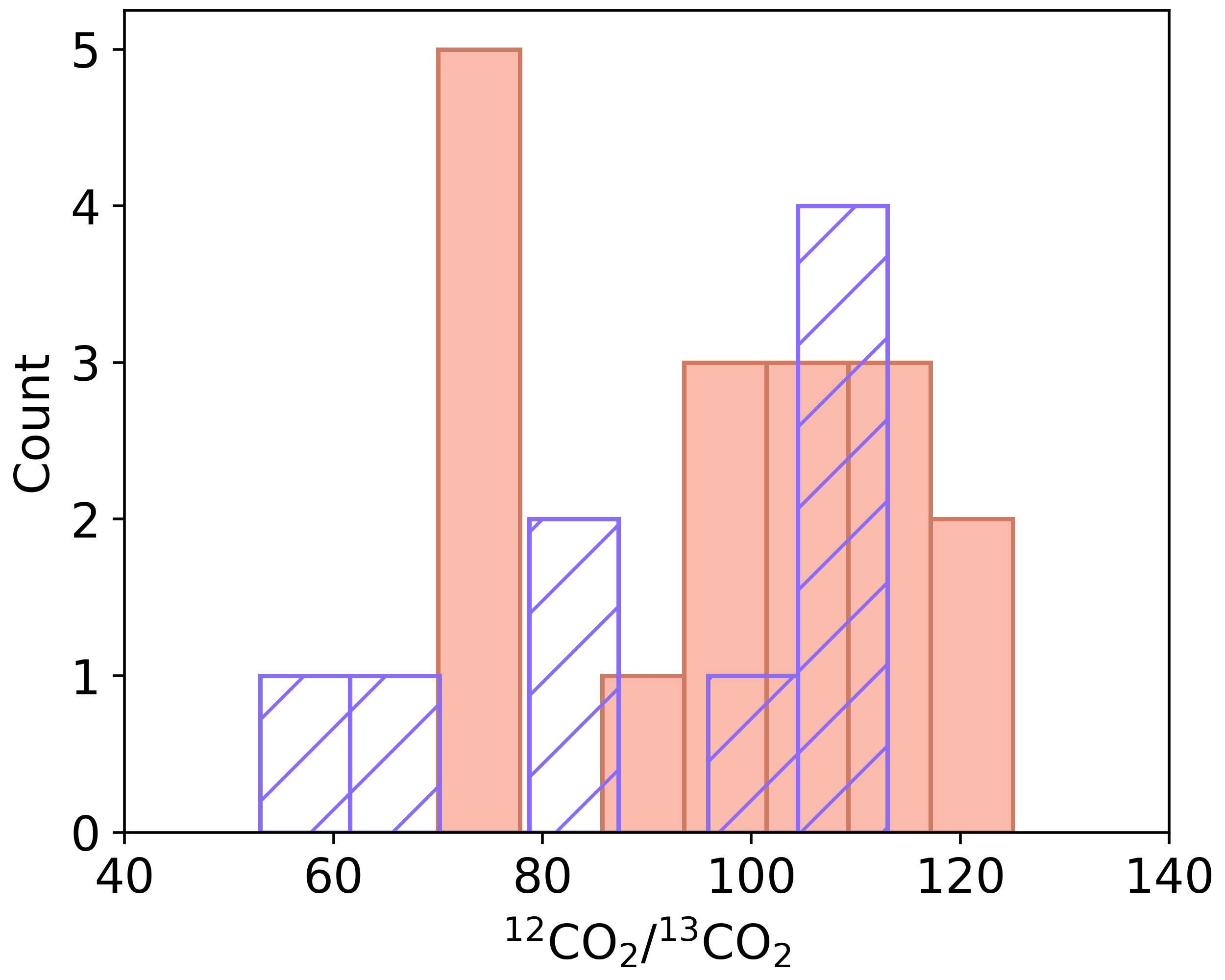}
    \caption{Distribution of the \ce{^{12}CO_2}/\ce{^{13}CO_2} isotope ratio. The orange filled bars show the distribution of the carbon isotope ratios determined for the JOYS+ sample. These ratios were extracted from the 15.2 \ce{\mu}m bending mode. The purple striped bars show the distribution of the carbon isotope ratio derived for the ISO high mass sources \citep{boogert1999iso}. }
    \label{fig:distribution}
\end{figure}

\subsection{Protostars: Solid state}
\label{subsec:co2solid}

\subsubsection{\ce{CO_2}}

The \ce{^{12}C}/\ce{^{13}C} ratios we derived from solid \ce{CO_2} for the JOYS+ sample are presented in Figure \ref{fig:ratios-CO2} along with isotope ratios derived in the solid state for the ISO-SWS study of young stellar objects \citep{boogert1999iso} and two highly extincted lines of sight \citep{mcclure2023ice} in the Cha I molecular cloud. We note that although previous studies used the uncorrected \citet{gerakines1995} band strengths, that the corrections have no affect on the relative band strengths and consequently the \ce{^{12}C}/\ce{^{13}C} ratios. For this reason, the ratios determined in these previous studies can be directly compared to the ratios derived in this work. The light error bars of the JOYS+ data points illustrate the uncertainties including the error on the band strengths, while the solid errorbars illustrate the uncertainties excluding the error on the band strengths. The errors in the band strengths account for the change in the band strength due to the temperature and ice mixture. The small error bars when the uncertainty on the band strengths are excluded demonstrate the great capabilities of the JWST and the high quality of these data. Note that previous studies generally do not include detailed analysis of the uncertainty in the band strengths in their errorbars.

The results show that the ratios obtained from the 15.2 \ce{\mu}m bands are clustered slightly above the ISM ratio of 68 \citep{boogert1999iso, milam2005}. This suggest that the ices in our sources are slightly deficient in \ce{^{13}C} in comparison to the local ISM.  In the cases where we detect all three \ce{^{12}CO_2} vibrational modes, we find consistent column densities and ratios (Figure \ref{fig:ratios-CO2}).  
This is further evidence that the 2.70 \ce{\mu}m feature primarily consists of the \ce{^{12}CO_2} combination mode. These consistent values also show that the 4.27 \ce{\mu}m stretching mode can be used to determine \ce{^{12}CO_2} column densities in cases where the other \ce{^{12}CO_2} bands are not available and where the stretching mode is not saturated.

When compared to other solid state studies of carbon isotope ratios, the \ce{^{12}CO_2}/\ce{^{13}CO_2} ratios measured in our sample are fairly consistent with the those found in the ISO-SWS high mass sources by \citet{boogert1999iso} (ranging 52 - 113). Similar to the JOYS+ sources, the authors report values that are scattered around the local ISM ratio though they also find sub-ISM values as low as 52. These sub-ISM ratios still agree with our values within the reported errorbars. Figure \ref{fig:distribution} illustrates the distribution of the \ce{^{12}CO_2}/\ce{^{13}CO_2} ratio for the JOYS+ sample and the ISO high mass sources \citep{boogert1999iso}. While a large fraction of the ISO MYSOs also display elevated ratios with respect to the ISM standard, the ratios of the JOYS+ protostars are in general higher with median values of  81, 75 and 100 for the 2.70 \ce{\mu}m combination mode, the 4.27 \ce{\mu}m stretching mode and the 15.2 \ce{\mu}m bending mode, respectively. These are similar to the mean values of 85 $\pm$ 23, 76 $\pm$ 12 and 97 $\pm$ 17 derived for these bands respectively. The agreement between the ratios derived from the different \ce{CO_2} vibrational modes of the JOYS+ protostars again shows the great capability of the JWST to accurately probe different lines of sight and observe the weak ice features in the 2 \ce{\mu}m region, the generally saturated strong stretching mode at 4.27 \ce{\mu}m as well as the bending mode at 15.2 \ce{\mu}m.

Our values are also consistent, albeit a bit higher, compared to the values reported in \citet{mcclure2023ice} for the two lines of sight in the Cha I molecular cloud (\ce{^{12}CO_2}/\ce{^{13}CO_2} $\sim$ 69 - 87). We note however that the \ce{^{12}CO_2} column densities derived for Cha I were extracted from the 4.27 \ce{\mu}m bands which are saturated in both lines of sight. Consequently, the \ce{^{12}CO_2} column density could be somewhat underestimated in those sources. Furthermore, the uncertainties reported in \citep{mcclure2023ice} do not include errors on the band strengths. Nonetheless, the consistency between the ratios derived from the protostellar stage and the dense cloud stage paint a harmonious picture that the \ce{CO_2} was likely inherited from its parent molecular cloud.

\subsubsection{CO}
\label{subsec:COdisc}

The \ce{^{12}CO}/\ce{^{13}CO} ratios derived from solid \ce{CO} for the JOYS+ sample are shown in Figure \ref{fig:ratios} along with isotope ratios derived from various studies both in the gas and solid state. Similar to \ce{CO_2}, the corrections on the CO band strengths \citep{gerakines1995} have no effect on the \ce{^{12}C}/\ce{^{13}C} ratio and values derived in previous solid state studies can therefore be directly compared to the findings in this work.  Overall our results show that the isotope ratios obtained from the CO 4.67 \ce{\mu}m stretching mode are considerably elevated with respect to the local ISM value. The ratios are also higher than the ratios previously derived for one MYSO \citep{boogert2002b} and one LYSO \citep{pontoppidan2003m}, both of which are very close to the ISM standard. This discrepancy is likely a result of our \ce{^{13}CO} column density being underestimated due to the weak \ce{^{13}CO} ice absorption feature and the strong CO gas line forest that dominates this spectral region. This is particularly the case for the two extreme outliers Per-emb22, Per-emb33 and Per-emb27 (shown only in Table \ref{Tab:CD-CO}), both of which exhibit strong gas phase lines and weak \ce{^{13}CO} ice bands (Figure \ref{fig:bands4micron3}). 

The ratio derived from the 2.35 \ce{\mu}m feature in Per-emb55-b is also elevated with respect to the local ISM standard. When compared to the ratios derived in molecular clouds however, our values are consistent with those found by \citet{mcclure2023ice} where they also report super-ISM \ce{^{12}CO}/\ce{^{13}CO} ratios (99-184) (Figure \ref{fig:ratios}).

\subsection{CO versus \ce{CO_2} isotope ratios}

\begin{figure*}[h!]
    \centering
     \includegraphics[width=1\hsize]{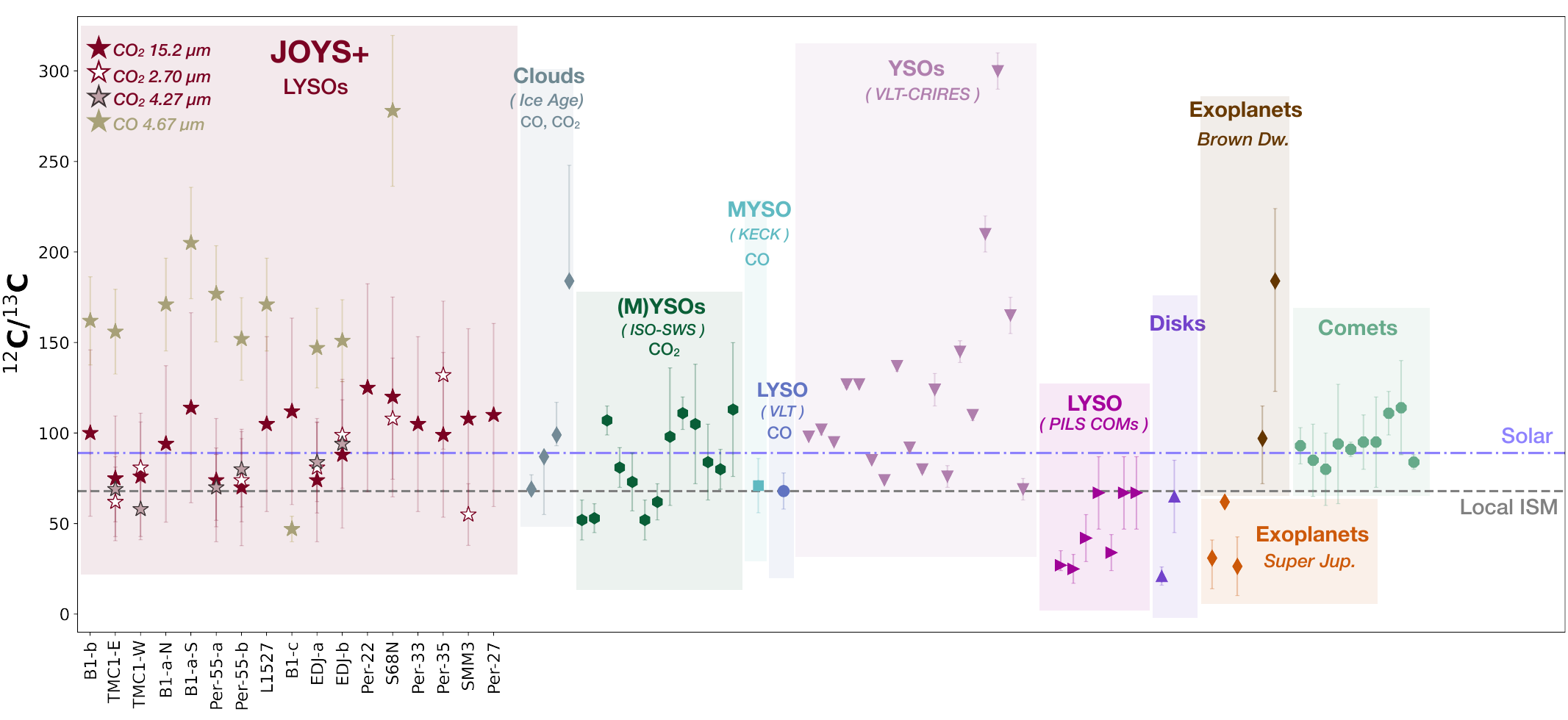}
    \caption{\ce{^{12}C}/\ce{^{13}C} for various astronomical environments. Red-filled, grey-filled and white-filled stars represent ratios derived from the \ce{^{12}CO_2} 15.2\ce{\mu}m, 4.27 \ce{\mu}m and 2.70 \ce{\mu}m bands respectively in this work. Yellow closed and open stars represent ratios derived from the \ce{^{12}CO} 4.67 
    \ce{\mu}m and 2.35 \ce{\mu}m bands respectively in this work. The errorbars on the JOYS+ data points include the error on the band strengths which is the major contributor to these uncertainties. Ice age \citep{mcclure2023ice}, \citep{boogert1999iso}, KECK-NIRSpec \citep{boogert2002b}, VLT \citep{pontoppidan2003m}, VLT-CRIRES \citep{smith2015}, PILS \citep{jorgensen2016, jorgensen2018}, Disks \citep{Yoshida2022a,bergin2024}, Exo-planets \citep{line2021, zhang2021,zhang2021b,ghandi2023,regt2024}, Comets \citep{bockelee2015,hassig2017}. dashed purple line shows the solar abundance and the dot-dashed gray line shows the local ISM ratio \citep{boogert1999iso}. Further details on the error analysis can be found in Appendix \ref{Appendix:A}}.
    \label{fig:ratios}
\end{figure*}

The ratios derived from CO and \ce{CO_2} from the JOYS+ sample are presented in Figure \ref{fig:ratios} along with gas phase and solid state ratios derived from various studies. Normally, the carbon isotope ratio of solid CO and \ce{CO_2} are expected to bear similarities since the formation route of \ce{CO_2} is tied to that of CO through equation \ref{eq:pathway1} \citep{ioppolo2011surface}:

\begin{equation}
    \ch{CO + OH -> HO-CO -> CO_2 + H } \label{eq:pathway1}
\end{equation}

In the ices of our sources however, we see that the \ce{^{12}CO}/\ce{^{13}CO} ratios are in general higher compared to the \ce{^{12}CO_2}/\ce{^{13}CO_2} ratios derived from the 2.70 \ce{\mu}m, 4.27 \ce{\mu}m and 15.2 \ce{\mu}m bands, though they still agree within the error bars for a number of our sources. As noted in Section \ref{subsec:COdisc} a reason for this deviation could be that the CO ratios are less accurate due to the weak \ce{^{13}CO} ice feature. Overall, we consider the \ce{CO_2}-derived isotope ratios to be more reliable given the multiple methods we used to quantify this ratio.

\subsection{Protostars: Gas phase}

In the gas-phase \citet{smith2015} reports significant \ce{^{12}CO}/\ce{^{13}CO} variability for several YSOs with an overall trend of elevated ratios relative to the local ISM standard (Figure \ref{fig:ratios}). One possible scenario they propose is gas-ice partitioning (Section \ref{subsubsec:partinioning}) with an additional mechanism for removing \ce{^{13}CO} from the solid state through the formation of larger complex organic molecules (COMs) from \ce{^{13}C}-rich CO ices. Consequently, the COMs will be enriched \ce{^{13}C} and exhibit low carbon isotope ratios, such as the sub-ISM values found in COMs by \citet{jorgensen2016,jorgensen2018}. We note however that the gas-phase values reported by \citet{smith2015} are similar to those derived in this study which suggests that the ices might have sublimated from the grains, delivering the carbon isotope ratio to the gas during the pre-stellar stage. 

Quantifying gas-phase and solid state isotope ratios in the same source was in fact proposed by \citet{charnley2004} as a way of tracing the origin of organic material. Therefore, observations of gas-phase CO isotopologues for the JOYS+ protostars will be crucial for enabling this method of posterior isotopic labeling and linking the solid state chemistry with the gas-phase chemistry in these systems. In the following section we will take a look at the different mechanisms that can affect the carbon isotope ratio in both the solid and gas phase in the early stages of star and planet formation. 

\subsection{Fractionation processes}
\label{subsec:fractioaning}

\subsubsection{Isotope exchange reactions}
\label{subsubsec:exchange}

Isotope exchange reactions are exothermic reactions that are efficient at low temperatures and that prompt preferential incorporation of \ce{^{13}C^+} in CO (Equation \ref{eq:exchange}) \citep{watson1976, langer1984, langerpenzias1993}. As a result, molecular \ce{^{13}CO} is enhanced with respect to \ce{^{12}CO} in the gas phase and this low \ce{^{12}CO}/\ce{^{13}CO} ratio is subsequently conveyed to the solid state upon CO freeze-out. The \ce{^{12}C^+} in contrast is enhanced in the gas phase leading to high \ce{^{12}C^+}/\ce{^{13}C^+} gas-phase ratios.

\begin{equation}
    \label{eq:exchange}
    \ch{^{13}C^+ +  ^{12}CO <-> ^{12}C^+ + ^{13}CO + $\Delta$ E = 34 K } 
\end{equation}

\subsubsection{Selective photodissociation}
\label{subsubsec:dissociation}

Selective photodissociation of \ce{^{13}CO} is a destruction mechanism that occurs because self-shielding of the less abundant \ce{^{13}CO} takes place at higher extinctions in comparison to \ce{^{12}CO} \citep{dieshoeckblack1988,visser2009}. This leaves a large portion of the \ce{^{13}CO} gas vulnerable to ultraviolet radiation. As a result, \ce{^{13}CO} is destroyed increasing the gaseous \ce{^{12}CO}/\ce{^{13}CO} ratio  and decreasing the \ce{^{12}C^+}/\ce{^{13}C^+} ratios as the molecules dissociate and the gas is enriched with \ce{^{13}C^+} and O atoms, which counters the effect of isotope selective photodissociation (Section \ref{subsubsec:exchange}). This high molecular \ce{^{12}CO}/\ce{^{13}CO} ratio translates to the solid state upon CO freeze-out producing ices that are depleted in \ce{^{13}C}. Although we expect that under dark cloud conditions most of the radiation is attenuated by the high density environment, thus limiting the effect of selective photodissociation, studies by \citet{furuya2018} have shown that turbulence can transfer selective photodissociated material to the dark inner regions of clouds.

\subsubsection{Gas-ice partitioning}
\label{subsubsec:partinioning}

A third mass-dependent mechanism was proposed by \citet{smith2015} where a small difference in the binding energies of \ce{^{12}CO} and the slightly heavier \ce{^{13}CO} (\ce{\Delta}\ce{E_{bind}} $\sim$ 10 K) results in \ce{^{12}CO} desorbing at a slightly lower temperature from the grains. Consequently, the \ce{^{12}CO}/\ce{^{13}CO} in the gas phase is enhanced while the ices become enriched with \ce{^{13}CO}, leading to low carbon isotope ratios in the solid state. It is worth nothing however that gas-ice partitioning is only effective during a very narrow temperature window \citep{smith2015}.

Of the mechanism discussed above, selective photodissociation could be the culprit of the \ce{^{13}C} deficiency observed in this sample though its efficiency under these conditions is open to question.


\subsection{Evolution of the \ce{^{12}C}/\ce{^{13}C} ratio}

As the stellar system matures, different process will begin to dominate, slowly eroding the initial carbon isotope ratio of the pristine material. In the following sections we will discuss the evolution of the carbon isotope ratio during the later epochs of star formation.

\subsubsection{Proto-planetary disks}

Studies of carbon isotope ratios in disks reveal that at this stage the conditions in the system are irrevocably changed and the initial carbon isotope imprint can be effectively erased in certain regions of the disk. \citet{Yoshida2022a} and \citet{bergin2024} for instance found a dichotomy in the TW Hya disk where two different carbon isotope ratios were measured for CO and \ce{C_2H} (21 $\pm$ 5 and 65 $\pm$ 20, respectively). In the same disk \cite{hily-blant2019} finds a ratio of 86 $\pm$ 4 from HCN rotational lines.

The variability of the carbon isotope ratio in protoplanetary disks is a consequence of the ratio being contingent on the location in which the molecules were formed. This is because different processes dominate in different regions of the disk. Photodissociation for instance will play a significant role in the upper layers of the disk, where the high ratio derived from \ce{C_2H} is found, while isotope exchange reactions are more efficient in the mid-plane and at larger radii where the temperatures are low and where the low value for CO is measured \citep{woods&willacy2009,visser2018, bergin2024}.

  


\subsubsection{Exoplanets}

The large variability of the carbon isotope abundance is also observed in the atmospheres of exo-planets where the formation mechanism of the planet plays a crucial role in its carbon isotope reservoir. 
Recent observational efforts by \citet{zhang2021,line2021, ghandi2023} for instance, revealed a significant enrichment of \ce{^{13}CO} in the atmospheres of hot Jupiters (31$\pm$17, 10.2-42.6 with 68\% confidence, and 62 $\pm$2, respectively). This \ce{^{13}C} enrichment is likely because the planet was formed beyond the CO snowline and accreted most of its material from \ce{^{13}C}-rich ices. These \ce{^{13}C}-rich ices are likely the result of fractionation processes such as isotope-exchange reactions and possibly gas-ice partitioning that must have occured after the protostellar stage.

\citet{zhang2021b,regt2024} in contrast measured very high \ce{^{12}C}/\ce{^{13}C} ratios (97 $\pm$ 15 and 184 $\pm$ 61, respectively) for brown dwarfs pointing to a formation route similar to that of stars where the object forms from cloud fragmentation and gravitational collapse and inherits most of its gaseous material from a parent cloud that is also deficient in \ce{^{13}C} and \ce{^{13}C}-ices as if found here. This shows that the carbon isotope ratio is in potentially a powerful tool for tracing the formation pathways of extrasolar planets.

\subsubsection{Comets and meteorites}

The chemical complexity of the infant planetary system is at last fossilized in the remains of its building blocks, the comets. Observations of \ce{C_2}, CN and HCN in numerous comets for instance showed that Solar system objects all have carbon isotope ratios similar to the Solar abundance ($\sim$ 89) with little variation between the ratios \citep{bockelee2015}. Similarly, \citet{alexander2007} finds ratios consistent with the Solar abundance from a large study of 75 chondrites. Finally, \cite{hassig2017} also derived a ratio of 84 $\pm$ 4 from \ce{CO_2} measurements in the coma of comet 67P/Churyumov-Gerasimenko. It is worth noting that the carbon isotope ratio found in comets and chondrites does bear some resemblance with the values derived in the protostellar stage. This could be an indication that while fractionation processes erode the initial carbon isotope imprint during the later stages, a fraction of the pristine material still survives this journey and is eventually incorporated into the planetary system.

\section{Conclusions}
\label{sec:5}

We have analyzed JWST NIRSpec and MIRI data of 17 Class 0/1 low mass protostars and determined the  \ce{^{12}CO}, \ce{^{13}CO}  \ce{^{12}CO_2} and \ce{^{13}CO_2} column densities and  \ce{^{12}C}/ \ce{^{13}C} isotope ratios from the \ce{^{12}CO_2} \ce{\nu_1} + \ce{\nu_2} and 2\ce{\nu_1} + \ce{\nu_2} combination modes (2.70 \ce{\mu}m and 2.77 \ce{\mu}m), the \ce{^{12}CO_2} \ce{\nu_3} stretching mode (4.27 \ce{\mu}m), the \ce{^{13}CO_2} \ce{\nu_3} stretching mode (4.39 \ce{\mu}m), the \ce{^{12}CO_2}  \ce{\nu_2} bending mode (15.2 \ce{\mu}m), the \ce{^{12}CO} 2-0 overtone mode (2.35 \ce{\mu}m), the \ce{^{12}CO} 1-0 stretching mode (4.67 \ce{\mu}m) and the \ce{^{13}CO} 1-0 stretching mode (4.78 \ce{\mu}m). The most significant finding is that the ratios are consistent, albeit slightly elevated, in comparison to the local ISM value. These results show that the ices leave the protostellar stage with high \ce{^{12}C}/\ce{^{13}C} ratios after which a series of fractionation processes during the later stages could modify the initial isotope abundance.
\begin{itemize}

    \item The absorption feature is observed at 2.70 \ce{\mu}m in 9 out 17 sources is assigned to the combination mode of \ce{^{12}CO_2}. We also detect the 2.77 \ce{\mu}m combination mode of \ce{^{12}CO_2} in two sources.
    \item We find consistent \ce{^{12}CO_2}/ \ce{^{13}CO_2} ratios between the 4.27 \ce{\mu}m, 15.2 \ce{\mu}m and the 2.70 \ce{\mu}m bands. Our values are slightly elevated with respect to the standard ISM value but consistent with carbon isotope ratios observed in other protostars and in dark molecular clouds. We observe variations of the \ce{^{12}CO_2}/ \ce{^{13}CO_2} ratio from source to source which could be pointing towards genuine differences in the chemical complexity of their protostellar envelopes.
    \item We report a detection of the 2.35 \ce{\mu}m \ce{\nu} 2-0 \ce{^{12}CO} overtone mode in one source.
    \item The \ce{^{12}CO}/\ce{^{13}CO} ratios derived from the 4.67 \ce{\mu}m band are elevated with respect to the local ISM ratio and the \ce{^{12}CO_2}/\ce{^{13}CO_2} ratios. The ratio derived from the overtone mode is a factor $\sim$ 1.6 lower compared to the ratio extracted from the 4.67 \ce{\mu}m band in the same source. 
   
\end{itemize}

Our findings show that fractionation processes begin to dominate after the protostellar stage. Future work should be focused on ice spatial mapping of close binary pairs and extended protostellar envelopes as well as obtaining observations of gas-phase CO isotopologue of the JOYS+ sample in order to compare solid state and gas-phase carbon isotope ratios.

\begin{acknowledgements}

Astrochemistry in Leiden is supported by the Netherlands Research School for Astronomy (NOVA), by funding from the European Re- search Council (ERC) under the European Union’s Horizon 2020 research and innovation programme (grant agreement No. 101019751 MOLDISK), and by the Dutch Research Council (NWO) grant 618.000.001. Support by the Danish National Research Foundation through the Center of Excellence “InterCat” (Grant agreement no.: DNRF150) is also acknowledged.

This work is based on observations made with the NASA/ESA/CSA James Webb Space Telescope. The data were obtained from the Mikulski Archive for Space Telescopes at the Space Telescope Science Institute, which is operated by the Association of Universities for Research in Astronomy, Inc., under NASA contract NAS 5-03127 for JWST.

The following National and International Funding Agencies funded and supported the MIRI development: NASA; ESA; Belgian Science Policy Office (BELSPO); Centre Nationale d’Etudes Spatiales (CNES); Danish National Space Centre; Deutsches Zentrum fur Luft- und Raumfahrt (DLR); Enterprise Ireland; Minis- terio De Economiá y Competividad; Netherlands Research School for Astron- omy (NOVA); Netherlands Organisation for Scientific Research (NWO); Sci- ence and Technology Facilities Council; Swiss Space Office; Swedish National Space Agency; and UK Space Agency.

NB thanks K. Pontoppidan for his contribution to the NIRSpec data reduction and helpful discussions. P.J.K. acknowledges support from the Science Foundation Ireland/Irish Research Council Pathway program under grant No. 21/PATH-S/9360. L.M. acknowledges the financial support of DAE and DST-SERB research grant (MTR/2021/000864) of the Government of India.

\end{acknowledgements}

\bibliographystyle{aa}
\bibliography{main}

\begin{appendix}

\twocolumn


\section{Error analysis}
\label{Appendix:A}

The uncertainties on the derived isotope ratios are mostly sensitive to the uncertainties in the band strengths. Minor contributors are also the error on the placement of the continuum and the noise level. To ultimately set constraints on the ratios we must take all these factors into account when calculating column densities. 

The ratio for the CO and \ce{CO_2} bands are calculated as follows:

\begin{equation}
    \label{eq:ratio}
    ratio = \frac{N_{^{12}CO_2}}{N_{^{13}CO_2}} = \frac{\frac{\int \tau d\nu_{^{12}CO_2}}{A_{^{12}CO_2}} }{  \frac{\int \tau d\nu_{^{13}CO_2}}{A_{^{13}CO_2}}  } = \frac{\int \tau d\nu_{^{12}CO_2}}{\int \tau d\nu_{^{13}CO_2}} \frac{A_{^{12}CO_2}}{A_{^{13}CO_2}},
\end{equation}

where $N_{^{12}CO_2}$ and $N_{^{13}CO_2}$ are the column densities of \ce{^{12}CO_2} and \ce{^{13}CO_2} respectively, $\int\tau d\nu_{^{12}CO_2}$ and $\int\tau d\nu_{^{13}CO_2}$ are the integrated optical depths and $A_{^{12}CO_2}$ and $A_{^{13}CO_2}$ are the band strengths of the \ce{^{12}CO_2} and \ce{^{13}CO_2} vibrational modes respectively.

From equation \ref{eq:ratio}, the uncertainty on the ratio is determined following the error propagation rules:

\begin{equation}
    \label{eq:errorratio}
    \frac{\sigma_{ratio}}{ratio} = \sqrt{ \left( \frac{ \sigma_{A_{^{12}CO_2}}}{A_{^{12}CO_2}} \right) ^2 + \left( \frac{ \sigma_{A_{^{13}CO_2}}}{A_{^{13}CO_2}} \right) ^2 + \left( \frac{ \sigma_{\tau d\nu_{^{12}CO_2}}}{\tau d\nu_{^{12}CO_2}} \right)^2 +  \left( \frac{ \sigma_{\tau d\nu_{^{13}CO_2}}}{\tau d\nu_{^{13}CO_2}} \right)^2     },
\end{equation}

where $\sigma_{\tau d\nu_{^{12}CO_2}}$ and $\sigma_{\tau d\nu_{^{13}CO_2}}$ are the uncertainties on the integrated optical depth for \ce{^{12}CO_2} and \ce{^{13}CO_2} respectively. 

The uncertainty on the integrated optical depth is influenced by the uncertainty in the placement of the continuum and the noise level and is determined as follows: 

\begin{equation}
    \left( \frac{ \sigma_{\tau d\nu_{^{12}CO_2}}}{\tau d\nu_{^{12}CO_2}} \right)^2_{^{12}CO_2} = \sqrt{\left( \frac{\sigma_{int. OD, cont}}{int. OD, cont} \right)^2_{^{12}CO_2} + \left( \frac{\sigma_{noise}}{int. OD, noise} \right)^2_{^{12}CO_2}},
\end{equation}

where $\sigma_{int. OD, cont}$ is the uncertainty due to the placement of the continuum, $int. OD, cont$ is the integrated optical depth with the continuum of choice, $\sigma_{noise}$ is the uncertainty due to the noise level and $int. OD, noise$ is the integrated optical depth. These parameters are calculated in a similar manner for \ce{^{13}CO_2}. 

\subsection{Continuum placement}

We determine the uncertainty on the placement of the continuum by fitting the bands with two different baselines and using the difference in integrated optical depth as the uncertainty. 

\subsection{Noise level}

The uncertainty due to the noise level is determined by calculating the rms in line free regions on flux scale ($\sigma_{f,\lambda}$) and propagating this to optical depth scale ($\sigma_{f,\tau}$) following the error propagation rules for logarithms:
 
\begin{equation}
    \label{eq:log}
    \sigma_{\tau,lambda} = \frac{\sigma_{f,\lambda}}{f_{cont,\lambda}},
\end{equation}

since the observed fluxes $F^{obs}_{\lambda}$ are converted to optical depth scale following equation (\ref{eq:OD}),

\begin{equation}
\label{eq:OD}
    \tau^{obs}_{\lambda} = -ln\biggl(\frac{F^{obs}_{\lambda}}{F^{cont}_{\lambda}}\biggl),
\end{equation}

where $F^{cont}_{\lambda}$ is the flux of the continuum.
\\
We take $F_{cont,\lambda}$ as constant since the uncertainty for that parameter is determined separately.

\begin{figure}[h!]
    \centering
     \includegraphics[width=1\hsize]{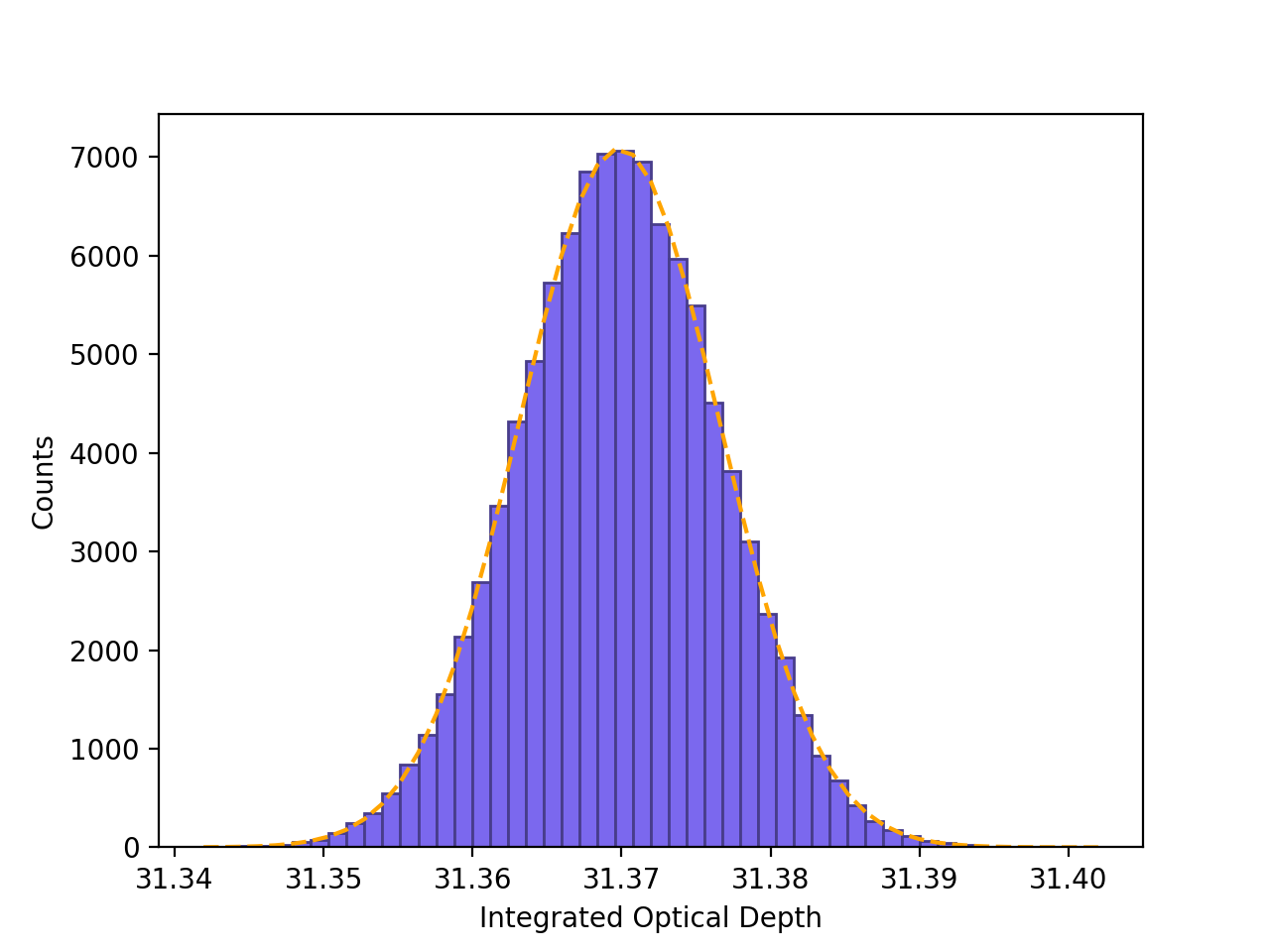}
    \caption{Sampling distrubition of the integrated optical depth. The dashed shows the Gaussian curve fitted to the distribution from which we extract $\sigma_{noise}$ and $int.OD,noise$.}
    \label{fig:histogram}
\end{figure}

We use $\sigma_{f,\lambda}$ to generate a noise level following a normal distribution and resample the spectrum with this added noise. We calculate the integrated optical depth for these resampled spectra which produces a sampling distribution that we can fit a Gaussian curve to (Figure \ref{fig:histogram}). From the Gaussian curve we can extract $\sigma_{noise}$ and $int.OD,noise$. This approach is similar to bootstrapping statistics and the contribution of this uncertainty is very small.

\subsection{Band strengths}

The column density for each vibrational mode is determined as follows:

\begin{equation}
\label{eq:CD}
    N = \frac{\int\tau d\nu}{A},
\end{equation}

where $A$ is the band strengths for of the corresponding vibrational mode.

As mentioned, we use the \citet{gerakines1995} and \citet{gerakines2005} band strength corrected by \citet{bouilloud2015} to calculate the column densities.

The contributing factors to the  uncertainty on the band strength for $A_{CO_2}$ are the placement of the continua (since we are considering relative errors), the uncertainty due to changes in the temperature of the ice and the uncertainty due to the ice mixtures. Since we know that the water component is a has a large contribution to the total band we took the uncertainties determined for the ice mixture \ce{CO_2}:\ce{H_2O} 1:24 binary ices. All values used in this work are reported in Table \ref{Tab:bandstrengths1}. 

For the error due to the continuum we assume an uncertainty of 5\% since \citet{gerakines1995} only reports uncertainties >5\% and no uncertainty is reported for this band strength. The uncertainty due to change in temperature is $\sim$ 10\%  \citep{gerakines1995}. Uncertainties for $A_{CO_2-H_2O}$ are reported in \citet{gerakines1995}.  

Taking all these assumption into consideration we determine the uncertainties on the band strengths as follows: 

\begin{equation}
\label{eq:error-bs}
   \sqrt{\sigma_{A,cont}^2 + \sigma_{A,temp}^2 + \sigma_{A, H_2O-CO_2}^2}   
\end{equation}

The uncertainties on the CO column densities are determined in a similar manner with following exceptions:
\begin{enumerate}
    \item We do not have mixture component for the error on the band strengths since we expect CO to be mainly in its pure form \cite{pontoppidan2003m}
    \item We do not have a temperature component for the error on the band strengths since CO disorbs above 40 K.
    \item We take a 10\% error on the band strength for both the \ce{^{12}CO} bands and the \ce{^{13}CO} to the placement of the continuum in laboratory spectra.
    \item The uncertainty on the integrated optical depth of the \ce{^{13}CO} ice feature between gas-subtracted spectra and non-gas subtracted spectra is $\sim$ 50 \%. This is not included in the error bars of the figures.  
\end{enumerate}

\clearpage

\onecolumn 
\section{Additional Figures}

\begin{center}
    
\begin{figure}[h!]
    \centering
     \includegraphics[width=0.6\hsize]{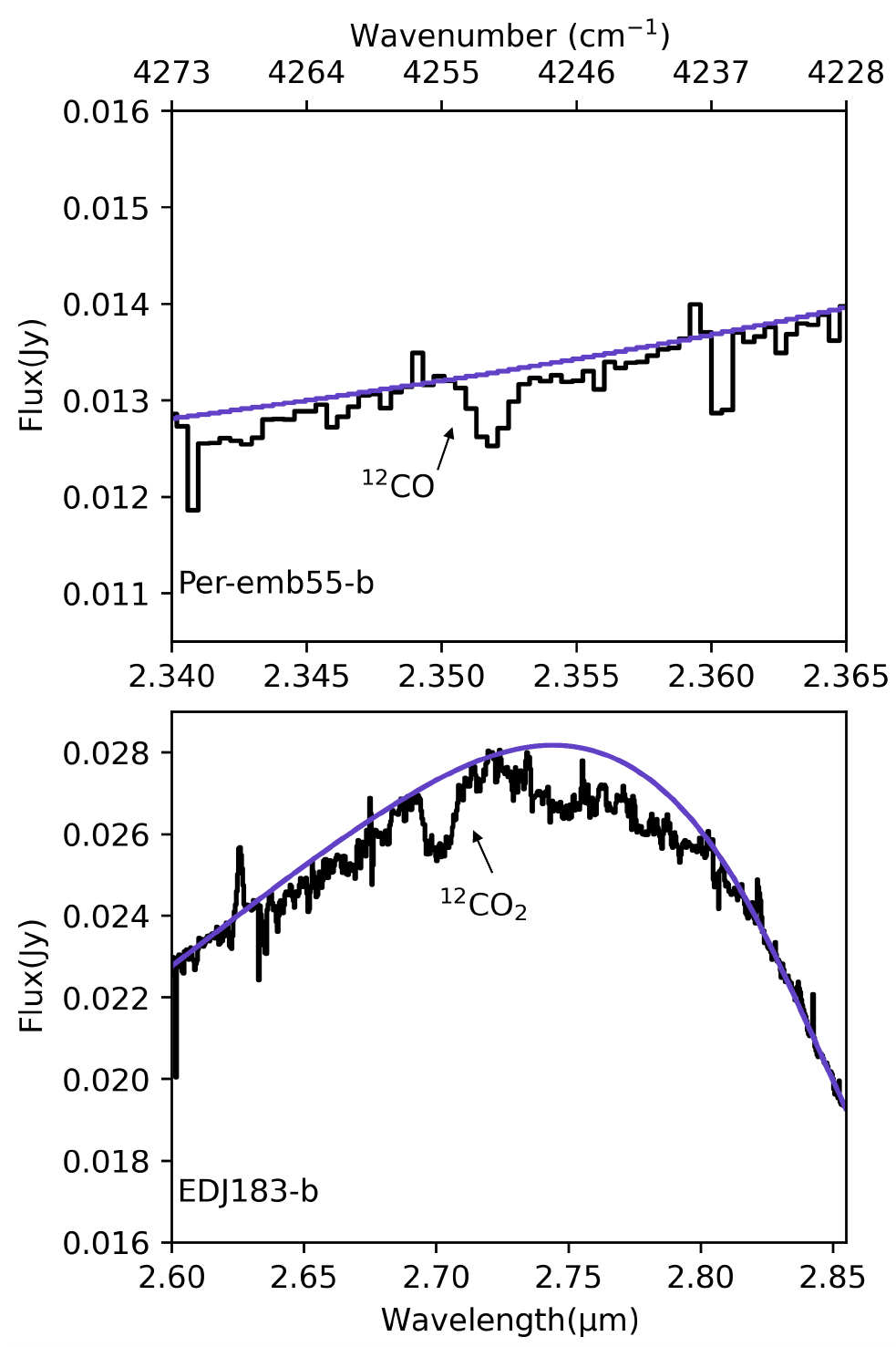}
    \caption{Continuum subtraction for the \ce{^{12}CO} overtone mode (left) and \ce{^{12}CO_2} combination mode (right). The purple line illustrates the local continuum drawn over each band.}
    \label{fig:cont2micron}
\end{figure}

\end{center}

\begin{figure*}[h!]
    \centering
     \includegraphics[width=0.9\hsize]{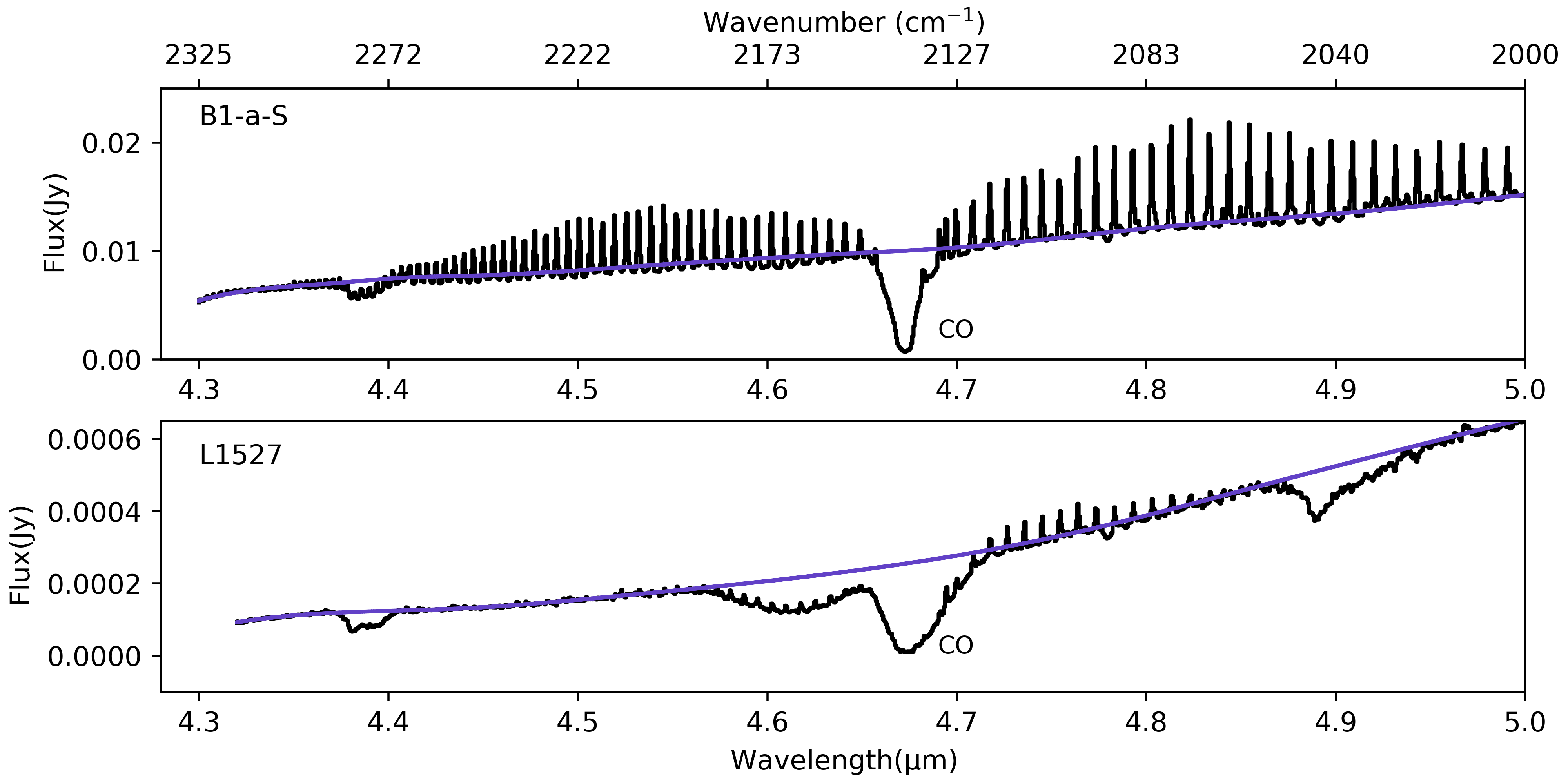}
    \caption{Continuum subtraction for the \ce{^{13}CO_2}, \ce{^{12}CO} and \ce{^{13}CO} stretching modes in the 4 \ce{\mu}m region for B1a-S (strong CO gas lines) and L1527 (weak CO gas lines). The purple line illustrates the local continuum placed over this region. }
    \label{fig:cont4micron}
\end{figure*}

\begin{figure*}[h!]
    \centering
     \includegraphics[width=0.8\hsize]{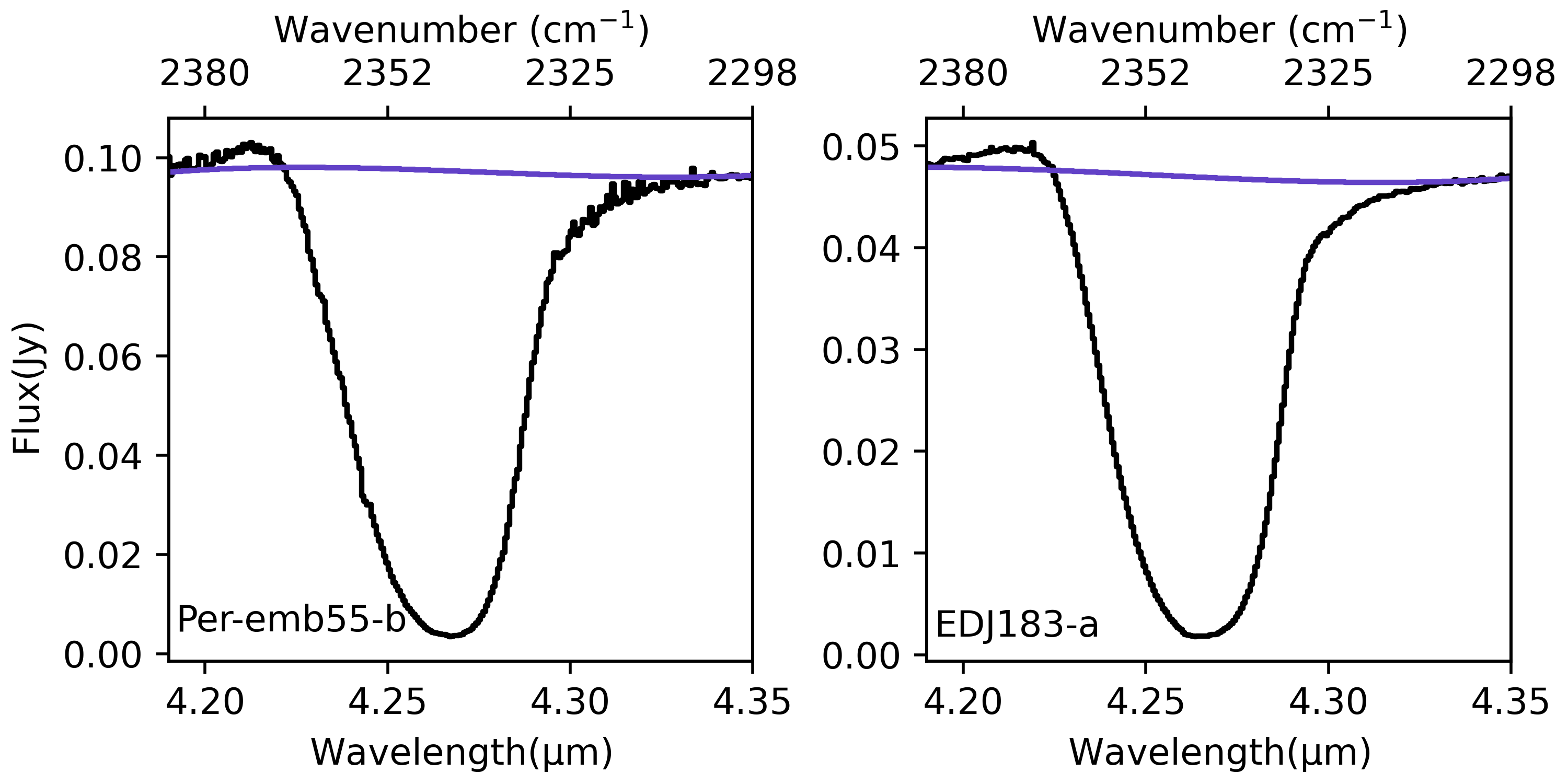}
    \caption{Continuum subtraction for the \ce{^{12}CO_2} 4.27 \ce{\mu}m stretching mode. The purple line illustrates the local continuum placed over the band. }
    \label{fig:continuum-co2-str}
\end{figure*}

In Figures \ref{fig:cont2micron}, \ref{fig:cont4micron} and \ref{fig:continuum-co2-str} we present the continuum subtractions for the 2 \ce{\mu}m region, the 4 \ce{\mu}m region and the \ce{^{12}CO_2} stretching mode at 4.27 \ce{\mu}m. 

\subsection{Continuum subtractions}

\FloatBarrier
\clearpage

\subsection{Band Overview}

\subsubsection{NIRSpec 4 \ce{\mu}m region }

\begin{figure*}[hbt!]
    \centering
     \includegraphics[width=0.9\hsize]{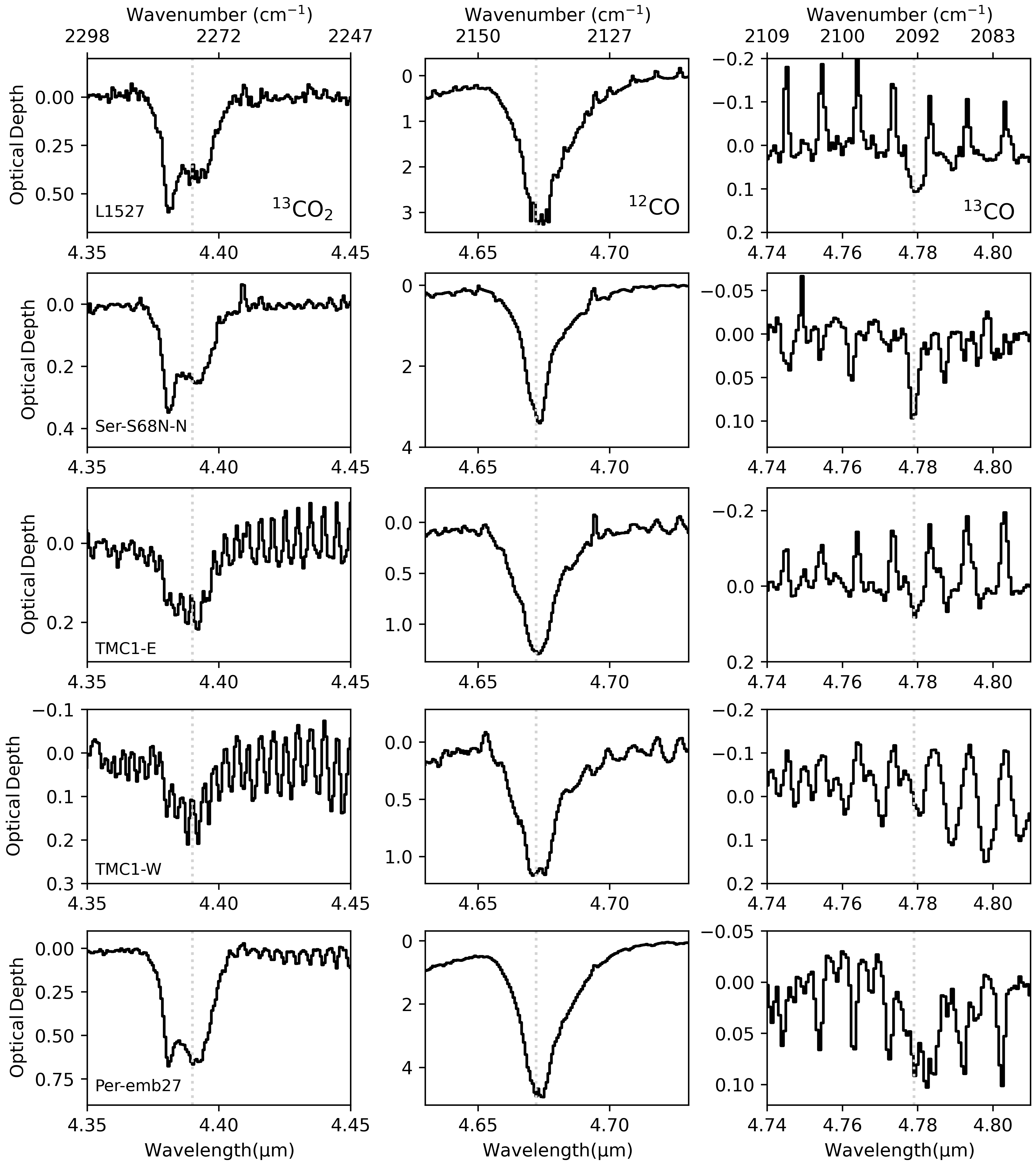}
    \caption{Overview of the \ce{^{13}CO_2} (left), \ce{^{12}CO} (middle) and \ce{^{13}CO} (right) stretching modes in the 4 \ce{\mu}m region.}
    \label{fig:bands4micron2}
\end{figure*}

\begin{figure*}[h!]
    \centering
     \includegraphics[width=0.9\hsize]{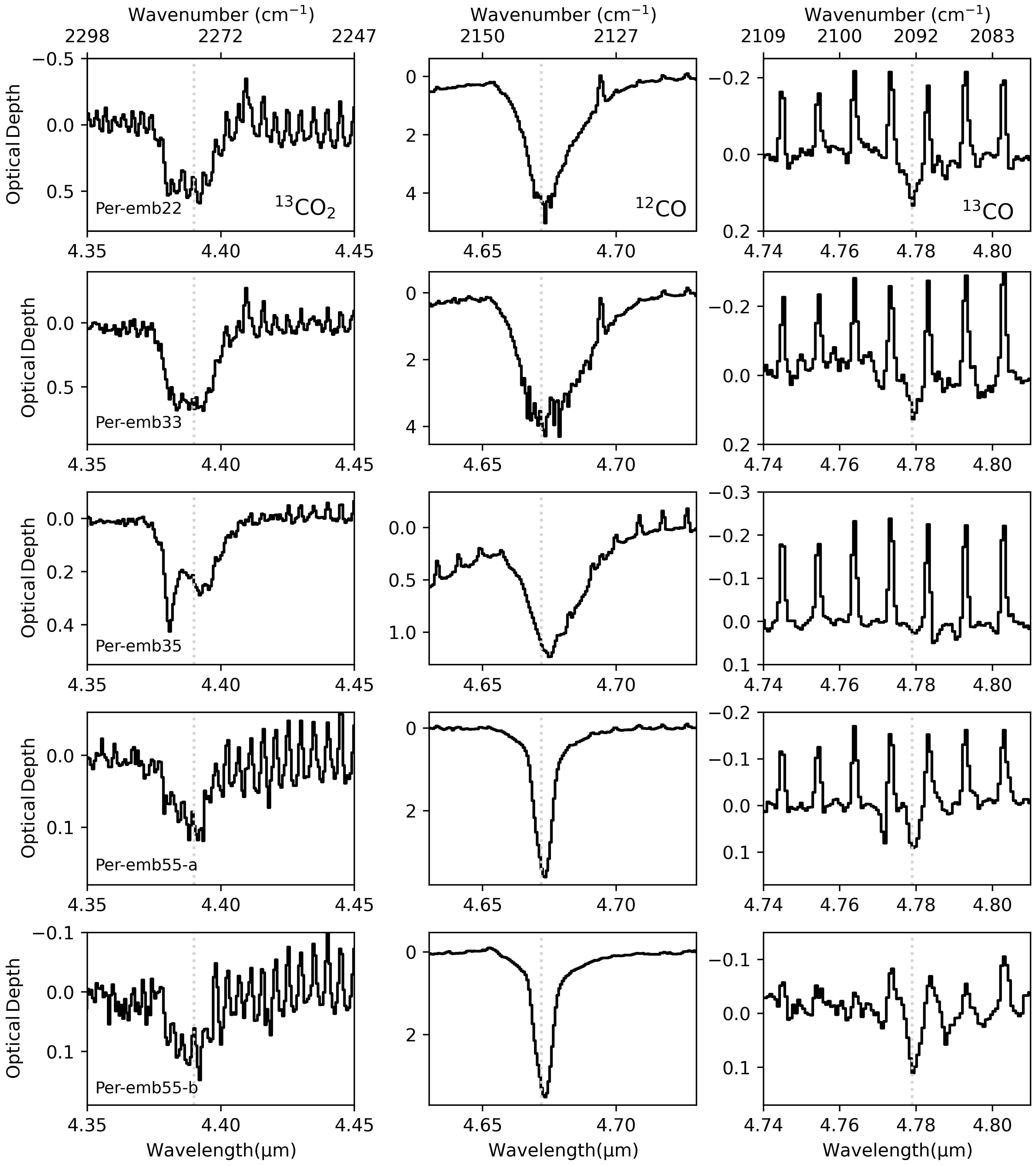}
    \caption{Overview of the \ce{^{13}CO_2} (left), \ce{^{12}CO} (middle) and \ce{^{13}CO} (right) stretching modes in the 4 \ce{\mu}m region.}
    \label{fig:bands4micron3}
\end{figure*}

\begin{figure*}[h!]
    \centering
     \includegraphics[width=0.9\hsize]{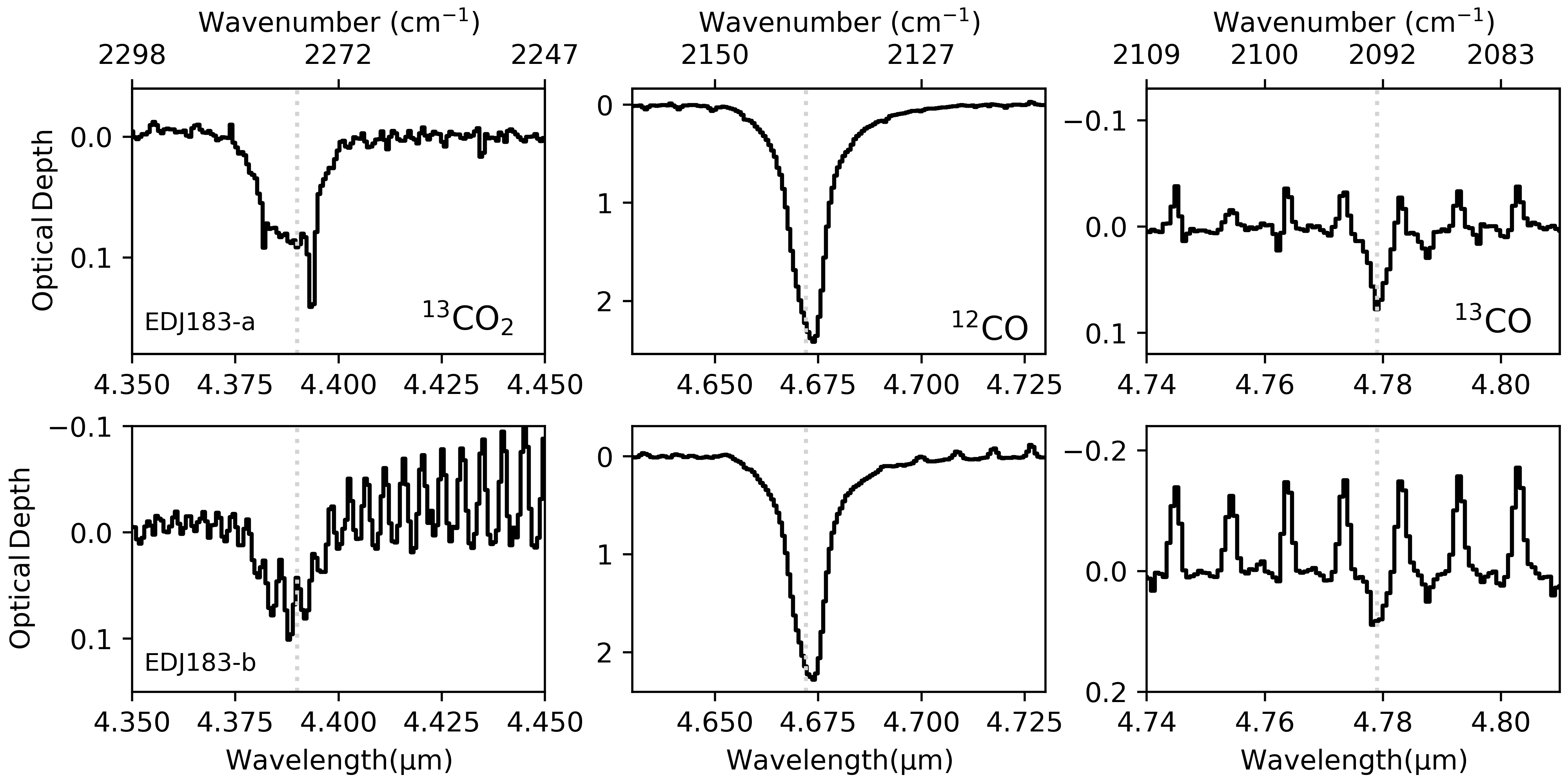}
    \caption{Overview of the \ce{^{13}CO_2} (left), \ce{^{12}CO} (middle) and \ce{^{13}CO} (right) stretching modes in the 4 \ce{\mu}m region.}
    \label{fig:bands4micron4}
\end{figure*}

\begin{figure*}[!ht]
    \centering
     \includegraphics[width=0.7\hsize]{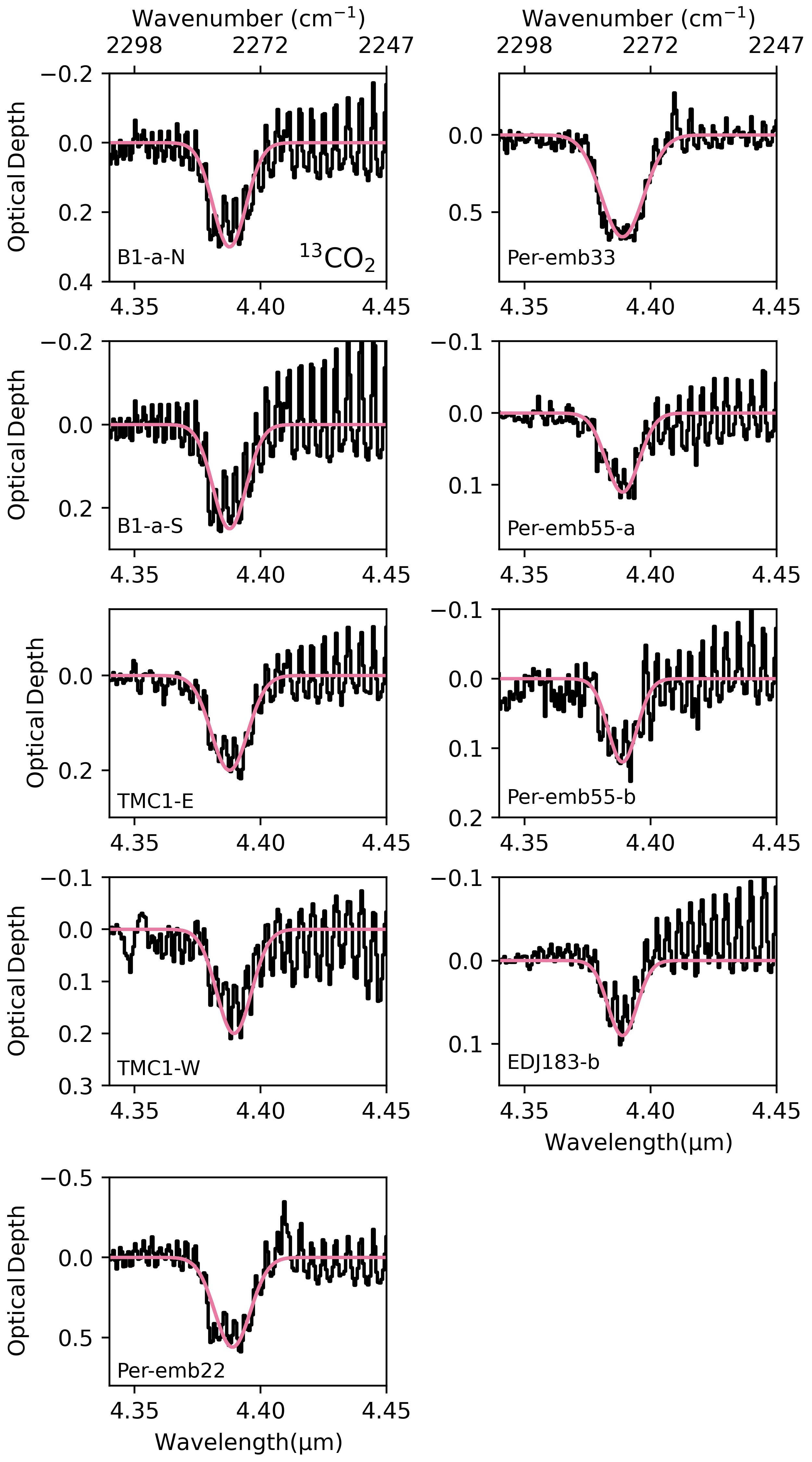}
    \caption{A zoom in of the \ce{^{13}CO_2} absorption features that suffer from gaseous CO line contamination. The Gaussian curves (pink) fitted to the bands illustrate the bounds used to determine the integrated optical depths. }
    \label{fig:bands13CO2}
\end{figure*}

\begin{figure*}[!ht]
    \centering
     \includegraphics[width=0.7\hsize]{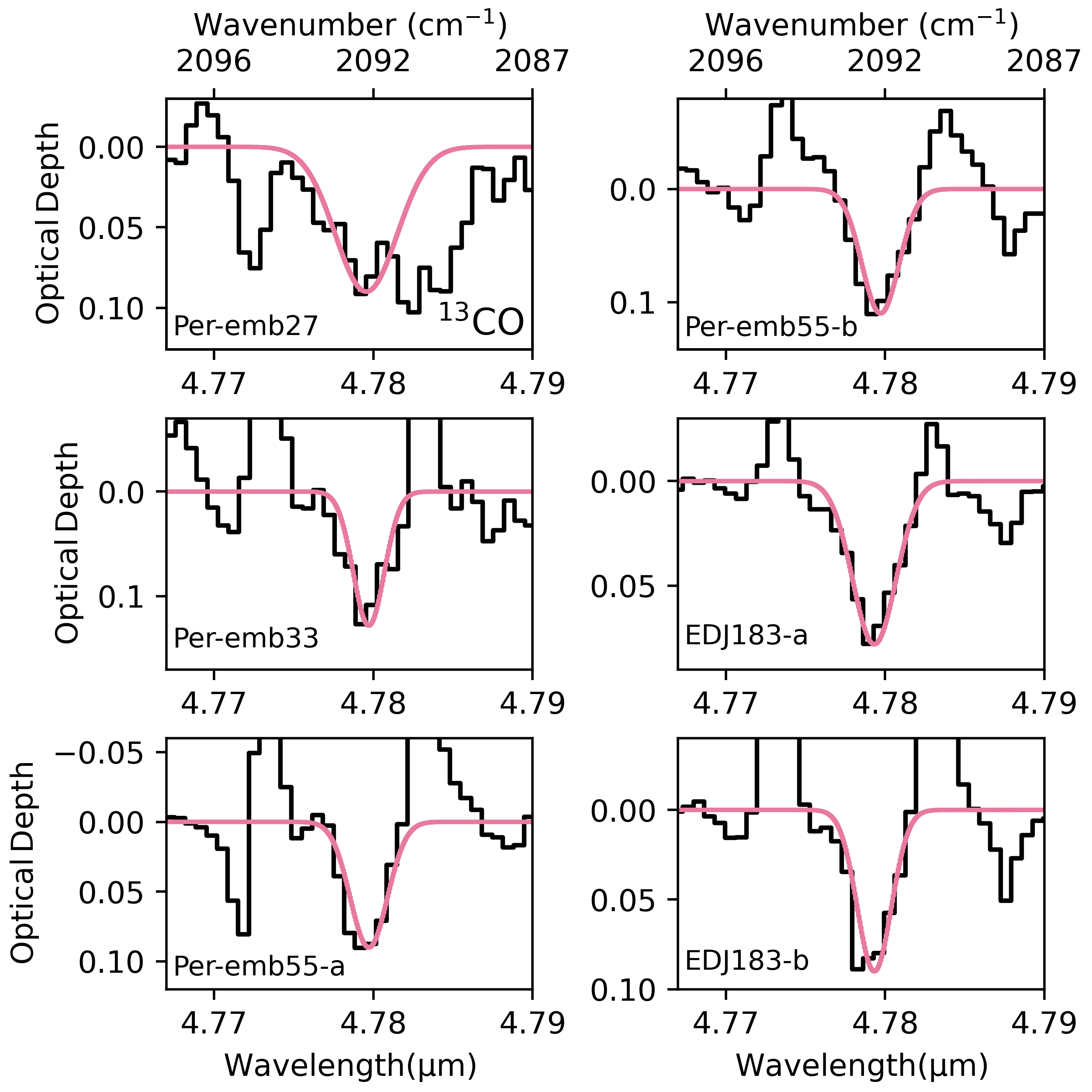}
    \caption{A zoom in of the \ce{^{13}CO} absorption features. The Gaussian curves (pink) fitted to the bands illustrate the bounds used to determine the integrated optical depths. }
    \label{fig:bands13CO-2}
\end{figure*}

\begin{figure*}[h!]
    \centering
     \includegraphics[width=0.9\hsize]{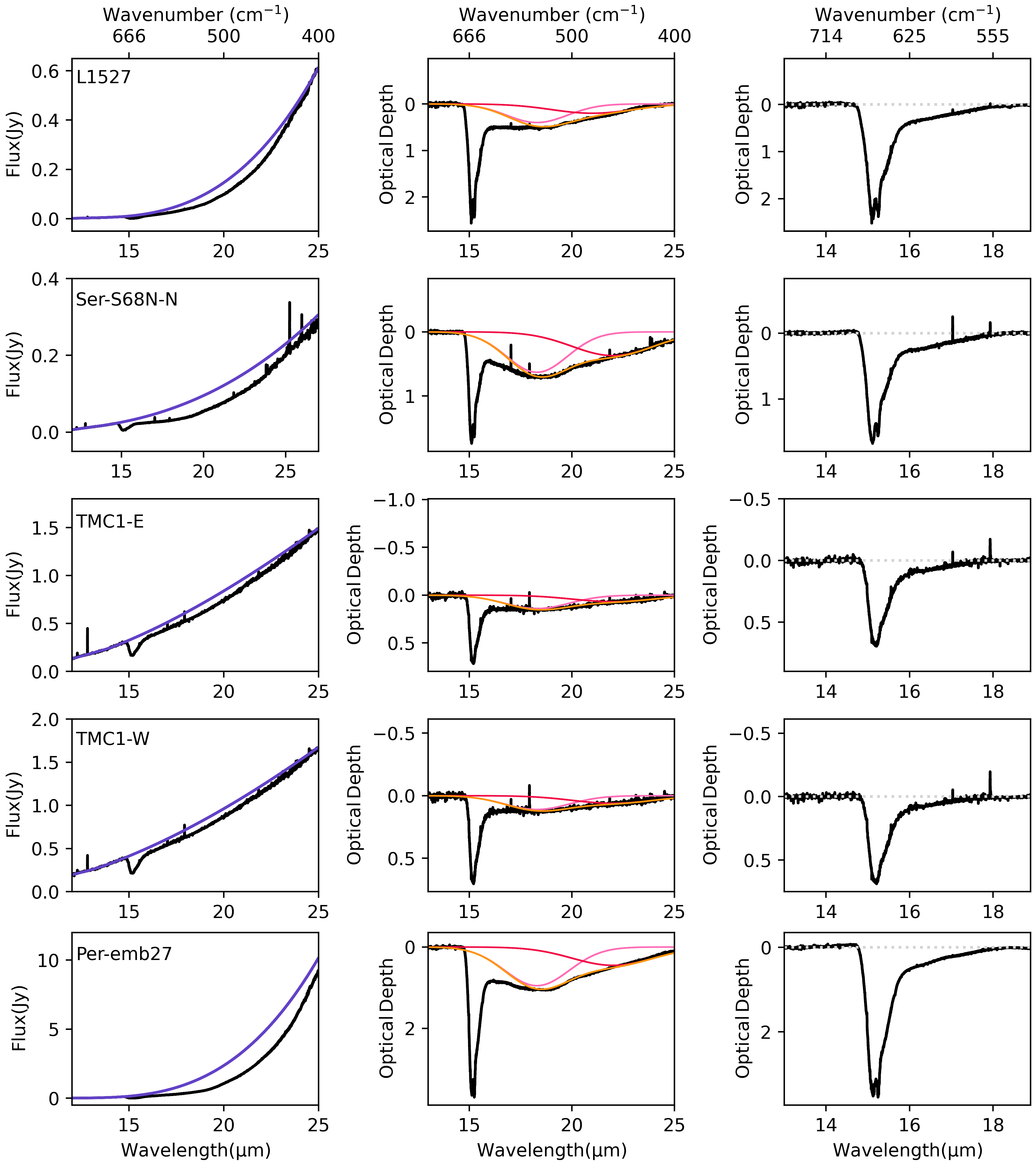}
    \caption{Continuum and silicate subtraction for the of the \ce{^{12}CO_2} bending mode in the 15 \ce{\mu}m region. The left panel shows the fitting of the continuum (purple line). The middle panel shows the two Gaussian profiles (red and pink lines) and the combined final fit (orange line) that simulates the silicate absorption feature at 18 \ce{\mu}m. For EDJ183-a and EDJ183-b an additional Gaussian profile (green line) was fitted to simulate the silicate feature in emission. The third panel shows the resulting spectra in optical depth scale.}
    \label{fig:cont15micron2}
\end{figure*}

\begin{figure*}[h!]
    \centering
     \includegraphics[width=0.9\hsize]{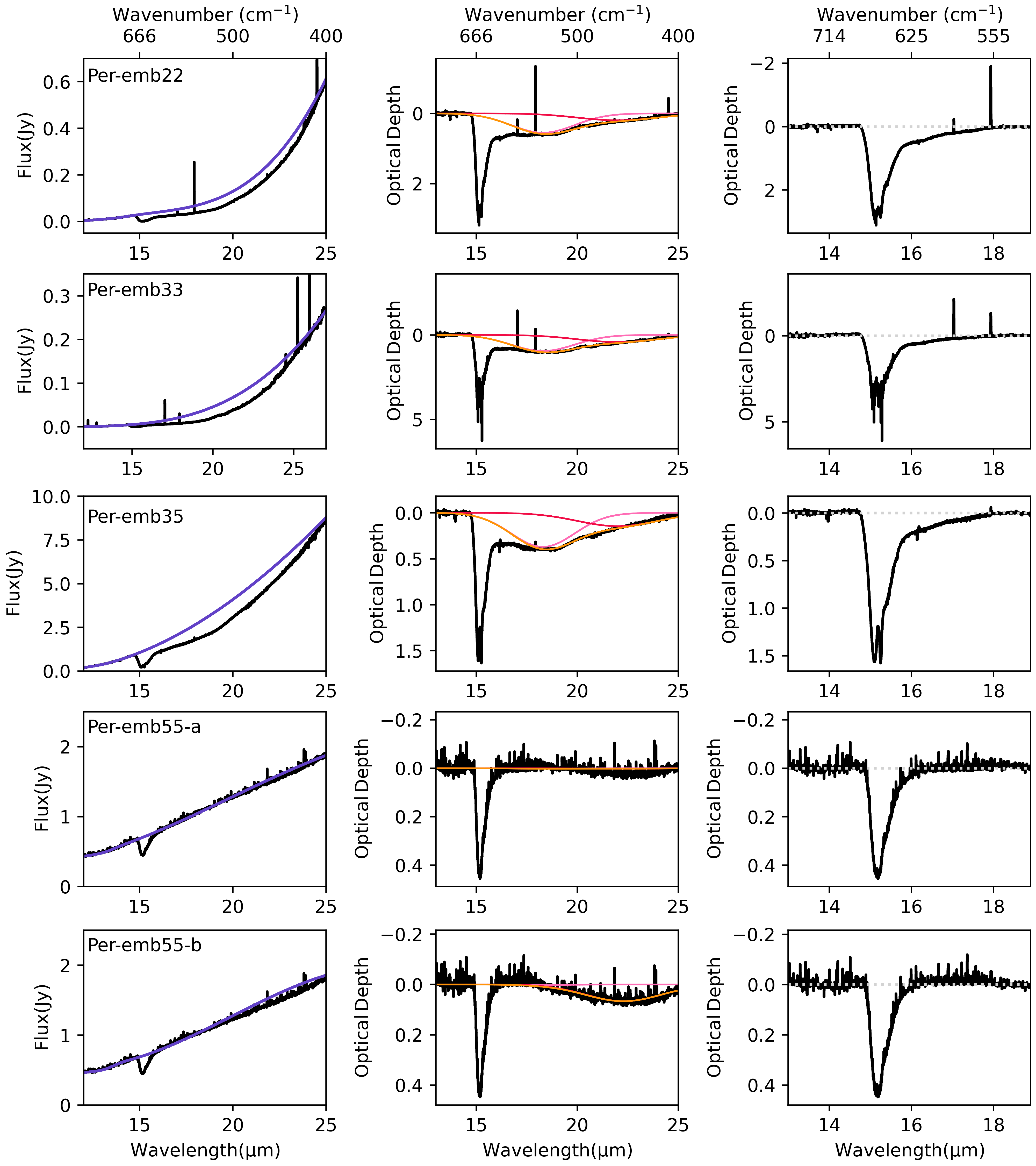}
    \caption{Continuum and silicate subtraction for the of the \ce{^{12}CO_2} bending mode in the 15 \ce{\mu}m region. The left panel shows the fitting of the continuum (purple line). The middle panel shows the two Gaussian profiles (red and pink lines) and the combined final fit (orange line) that simulates the silicate absorption feature at 18 \ce{\mu}m. For EDJ183-a and EDJ183-b an additional Gaussian profile (green line) was fitted to simulate the silicate feature in emission. The third panel shows the resulting spectra in optical depth scale.}
    \label{fig:cont15micron3}
\end{figure*}

\begin{figure*}[h!]
    \centering
     \includegraphics[width=0.9\hsize]{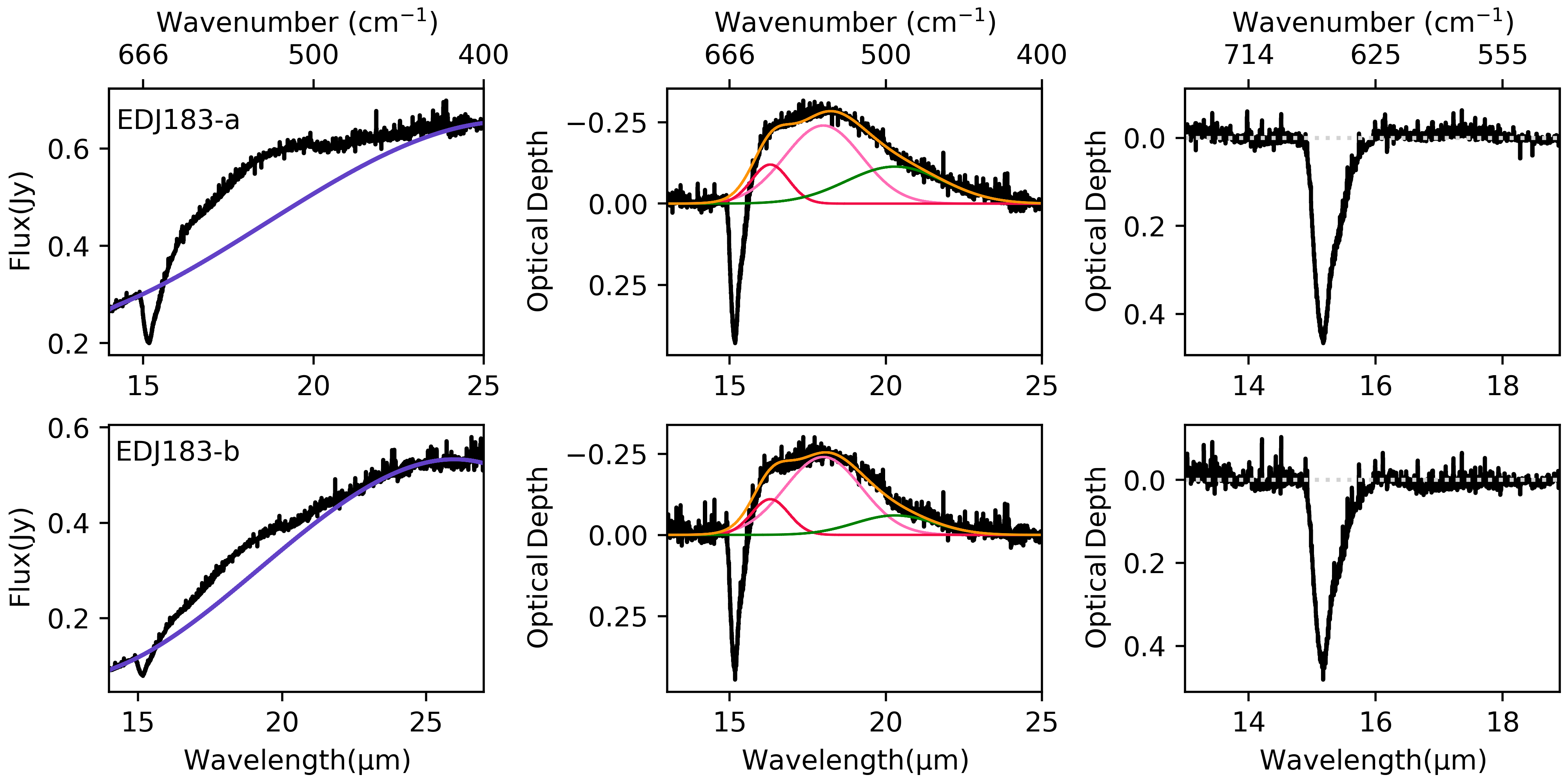}
    \caption{Continuum and silicate subtraction for the of the \ce{^{12}CO_2} bending mode in the 15 \ce{\mu}m region. The left panel shows the fitting of the continuum (purple line). The middle panel shows the two Gaussian profiles (red and pink lines) and the combined final fit (orange line) that simulates the silicate absorption feature at 18 \ce{\mu}m. For EDJ183-a and EDJ183-b an additional Gaussian profile (green line) was fitted to simulate the silicate feature in emission. The third panel shows the resulting spectra in optical depth scale.}
    \label{fig:cont15micron4}
\end{figure*}

In Figures \ref{fig:bands4micron2}, \ref{fig:bands4micron3}, \ref{fig:bands4micron4}, \ref{fig:bands13CO2}, \ref{fig:bands13CO-2} \ref{fig:cont15micron2}, \ref{fig:cont15micron3} and \ref{fig:cont15micron4} we present an overview of the \ce{CO_2} and CO ice features in the 4 \ce{\mu}m and 15 \ce{\mu}m region. 

\FloatBarrier

\clearpage
\subsubsection{MIRI 15 \ce{\mu}m region }
\label{app:MIRI-continuum}

\FloatBarrier

\subsection{Additional features}

\begin{figure}[!ht]
    \centering
     \includegraphics[width=0.6\hsize]{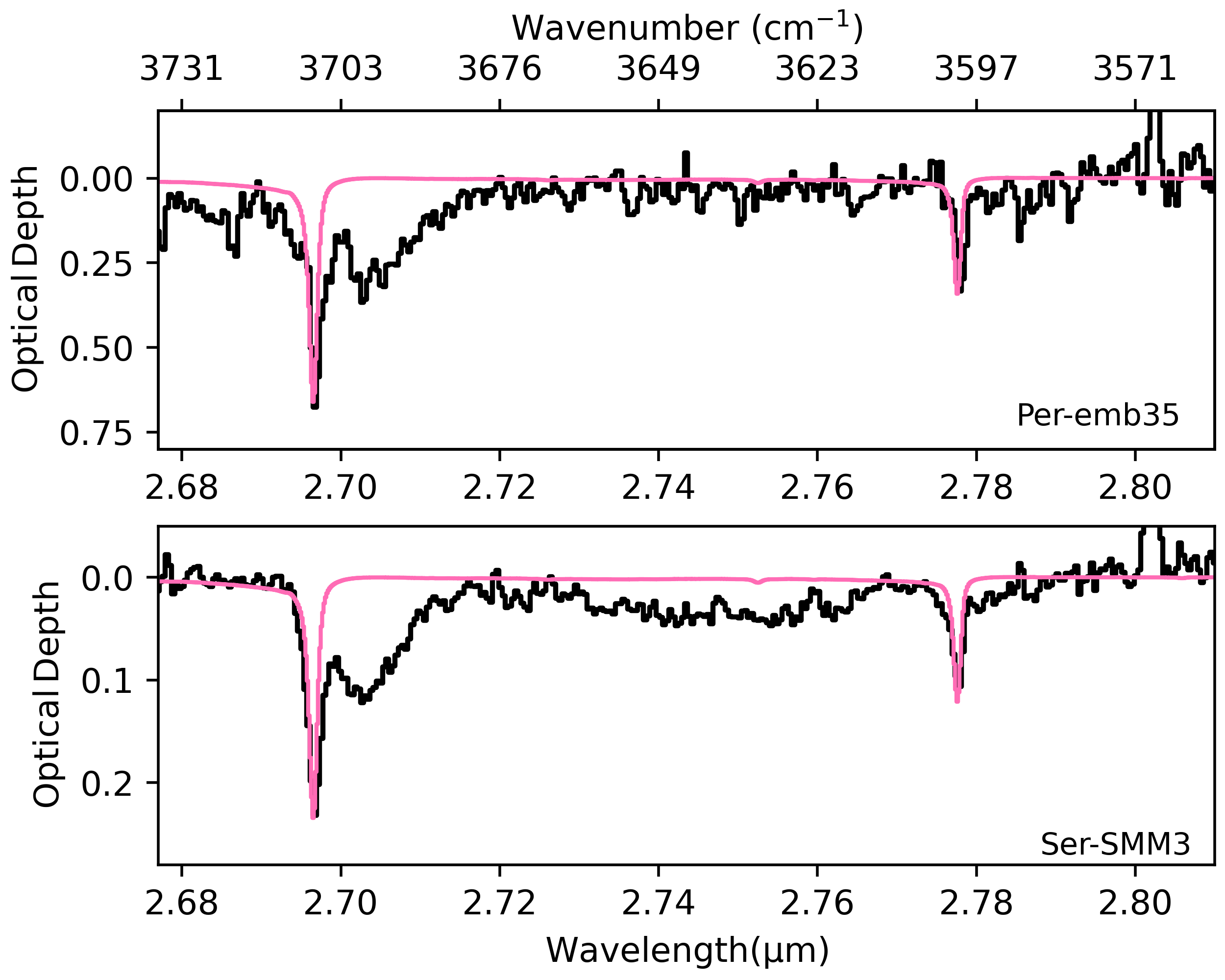}
    \caption{Laboratory fittings of the 2.70 and 2.77 combination modes. The spectrum of pure \ce{CO_2} at 80 K \citep{ehrenfreund1997infrared} is plotted over the two absorption feature in pink.}
    \label{fig:secondcombi}
\end{figure}

\begin{figure*}[h!]
    \centering
     \includegraphics[width=0.9\hsize]{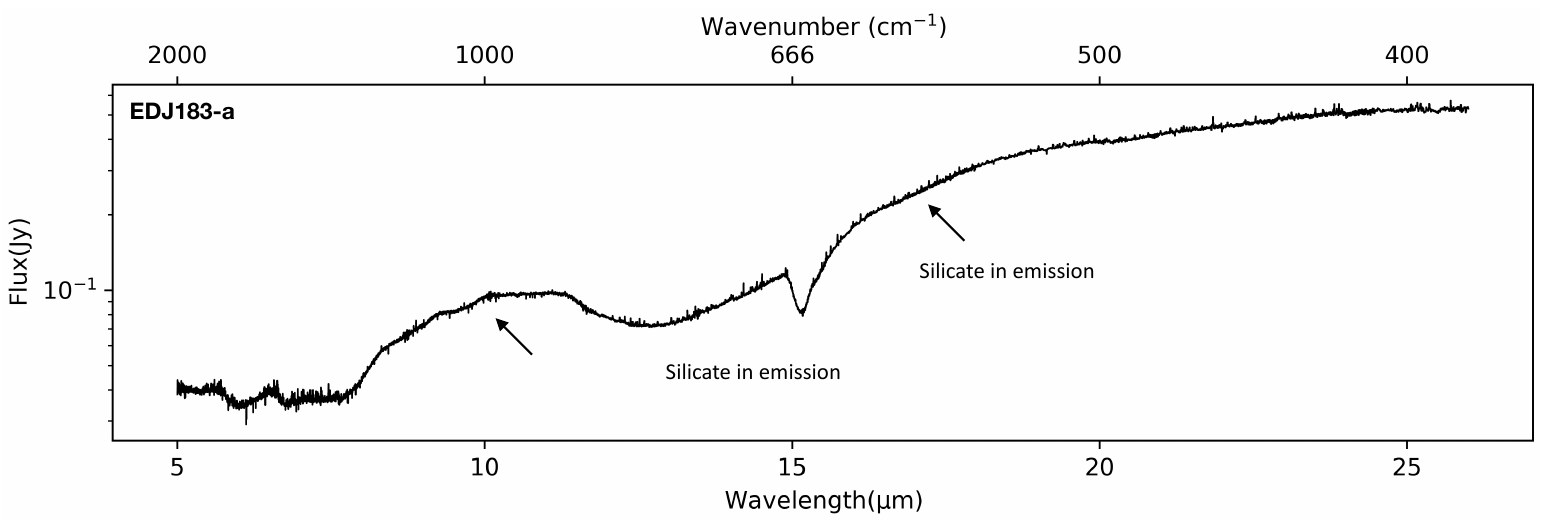}
    \caption{MIRI spectrum of EDJ183-a showing the 9 and 18 \ce{\mu}m silicate features in emission. The 18 \ce{\mu}m silicate bending mode is likely responsible for the distortion of the long-wavelength wing of the \ce{CO_2} absorption band at 15.2 \ce{\mu}m.}
    \label{fig:silicate-emission}
\end{figure*}

\begin{figure*}[h!]
    \centering
     \includegraphics[width=0.6\hsize]{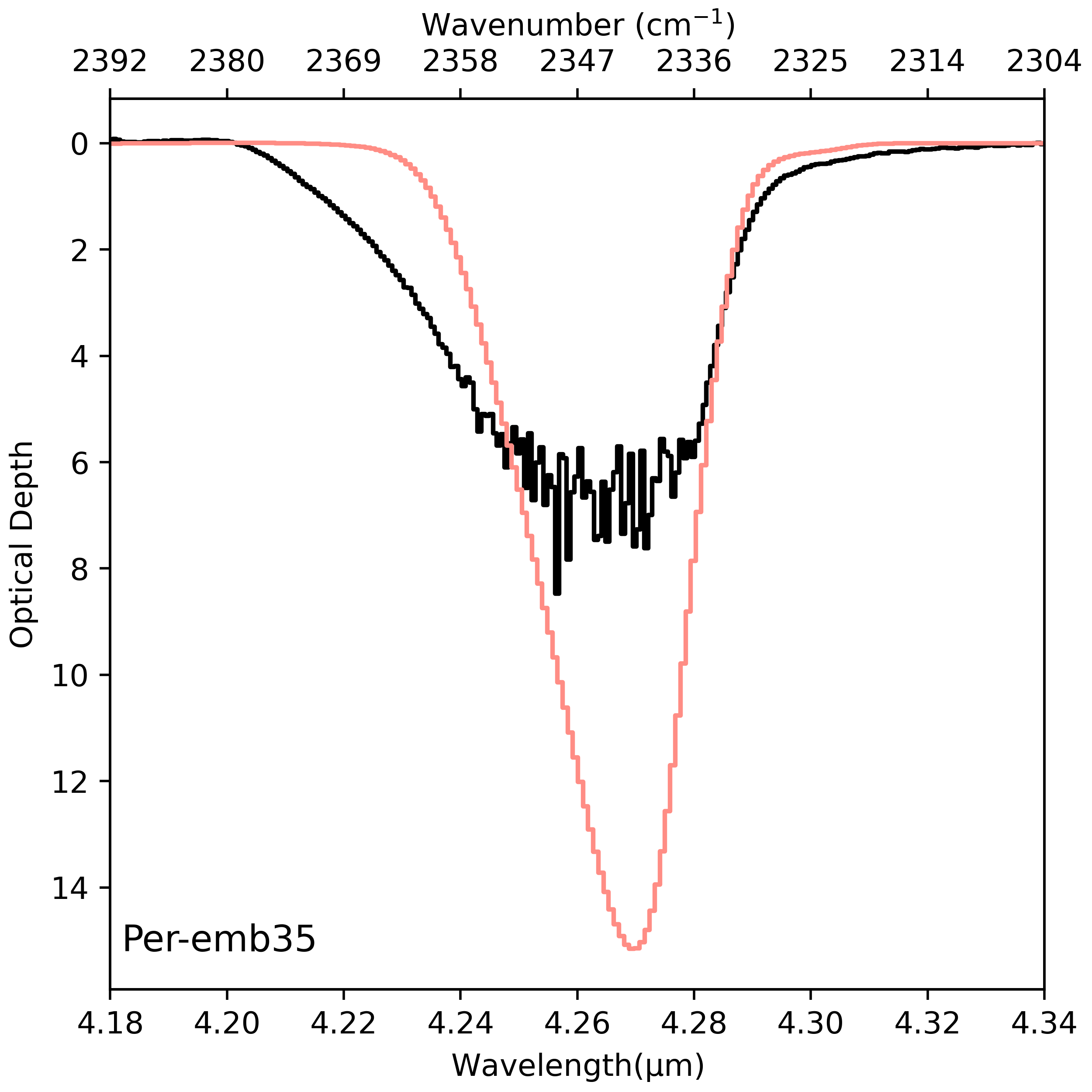}
    \caption{The 4.27 \ce{\mu}m band of Per-emb35. Plotted over the band is the \ce{CO_2}:\ce{H_2O} 1:10 spectrum at 10 K \citep{ehrenfreund1999laboratory}.}
    \label{fig:per35-str}
\end{figure*}

\begin{figure*}[h!]
    \centering
     \includegraphics[width=1\hsize]{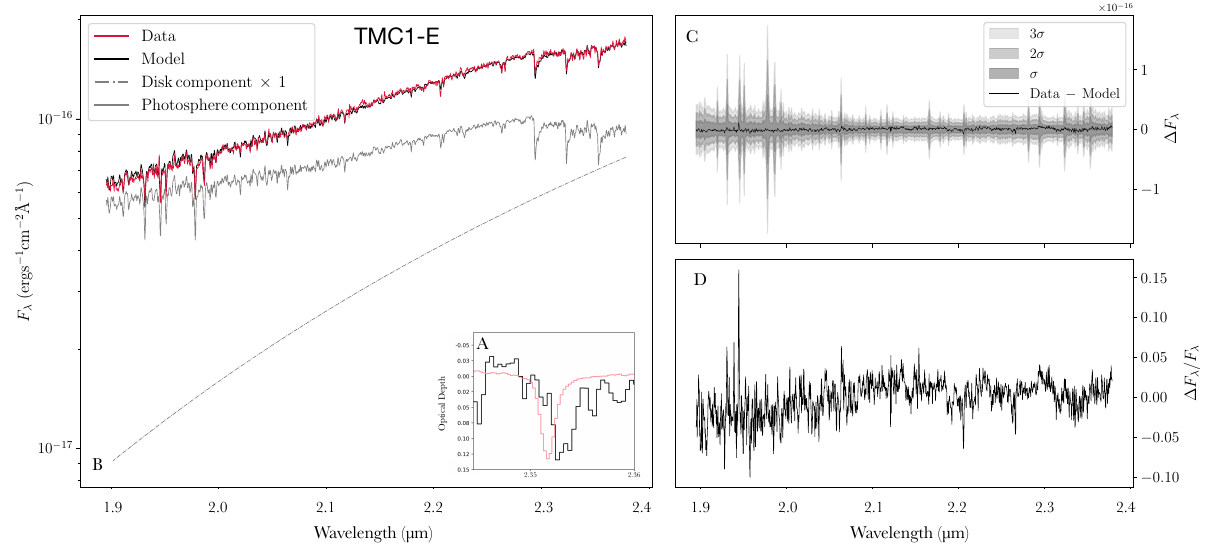}
    \caption{Photospheric model fittings for TMC1-E using the Starfish MCMC exploration \citep[Le Gouellec in prep.]{legouellec2024}. Panel A shows the 'red-shifted` 2.35 \ce{\mu}m absorption feature with the spectrum of pure CO ice plotted over the feature (pink). Panel B shows the full 1.9 - 2.45 \ce{\mu}m spectrum (red) with the best fit model plotted over the data (black). The grey dot-dashed line shows the disk component of the model while the grey solid line shows the photosphere component of the model. In panel C we show the residuals and the diagonal of the covariance matrix as \ce{\sigma} contours. Finally, in panel D we show the relative errors (residuals to data flux ratio).}
    \label{fig:tmc1-a-model}
\end{figure*}

Figure \ref{fig:secondcombi} shows the 2.70 \ce{\mu}m and 2.77 \ce{\mu}m combination modes of \ce{CO_2} with the laboratory spectrum of pure \ce{CO_2} fitted to the data. Figure \ref{fig:silicate-emission} shows the silicate bands in emission in the source EDJ183-a. In Figure \ref{fig:per35-str} we show the \ce{^{12}CO_2} 4.27 \ce{\mu}m band of Per-emb35 and finally in Figure \ref{fig:tmc1-a-model} we show the photospheric model fittings for TMC1-E.

\FloatBarrier

\section{Source coordinates}

In Table \ref{Tab:coordinates} we present the coordinates of the apertures used to extract the spectra. 

\label{sec:coordinates}

\begin{center}
\begin{table}[hbt!]
\caption{Coordinates of extracted spectra.}
\small
\centering
\begin{tabular}{lcccc}
\hline \hline
Source &  RA NIRSpec & Dec NIRSpec & RA MIRI & Dec MIRI 
\\     
\hline

B1-a-N &	3:33:16.693 &	31:07:55.274 &	3:33:16.692 &	31:07:55.286 \\

B1-a-S &	3:33:16.677	& 31:07:54.969	& 3:33:16.681 &	31:07:54.897	\\	

B1-b	& 3:33:20.334 & 31:07:21.634	& 3:33:20.354 & 31:07:21.353 \\																
B1-c	& 3:33:17.893 & 31:09:31.900 & 3:33:17.893 & 31:09:31.847 \\																														
L1547 &	4:39:53.875 & 26:03:09.5940	& 4:39:53.836 & 	26:03:10.153 \\	

Per-emb22 &	3:25:22.351 &	30:45:13.186 &	3:25:22.349	& 30:45:13.111 \\

Per-emb27 &	3:28:55.575	& 31:14:36.900 &	3:28:55.573	&31:14:36.764 \\

Per-emb33 &	3:25:36.308	&30:45:15.018 &	3:25:36.304	& 30:45:14.958 \\

Per35 &	3:28:37.093 &	31:13:30.837	& 3:28:37.100 &	31:13:30.657 \\

Per55-a	& 3:44:43.292 & 	32:01:31.428 & 3:44:43.293 &	32:01:31.187 \\

Per55-b & 3:44:43.324 &	32:01:31.819 &	3:44:43.325	& 32:01:31.592 \\

EDJ183-a & 3:28:59.301 &	31:15:48.473	& 3:28:59.301 &	31:15:48.314 \\

EDJ183-b &	3:28:59.379	& 31:15:48.465 & 3:28:59.381 &	31:15:48.314 \\

Ser-S68N-N &	18:29:48.134 &	1:16:44.574	& 18:29:48.133 &	1:16:44.489 \\

Ser-SMM3 & 18:29:59.307	& 1:14:00.680 & 18:29:59.317 & 	1:14:00.732 \\

TMC1-E &	4:41:12.730	& 25:46:34.524 &	4:41:12.735 &	25:46:34.715 \\

TMC1-W	& 4:41:12.684 &	25:46:34.518 &	4:41:12.686 &	25:46:34.730 \\

\hline
\end{tabular}
\label{Tab:coordinates}
\begin{tablenotes}\footnotesize
\item{} 
\end{tablenotes}
\end{table}  
\end{center}

\end{appendix}

\end{document}